\documentstyle[preprint,aps]{revtex}
\input epsf
\tighten

\newcommand{\bc}{\begin{center}}
\newcommand{\ec}{\end{center}}
\newcommand{\beqn}{\begin{equation}}
\newcommand{\eeqn}{\end{equation}}
\newcommand{\barr}{\begin{eqnarray}}
\newcommand{\earr}{\end{eqnarray}}
\def\etal {{\it et al}. }
\def\eg {{\it e.g}. }
\def\ie {{\it i.e}. }

\def\rvac {|0\rangle}
\def\lvac {\langle 0|} 

\def\simge{
    \mathrel{\rlap{\raise 0.511ex
        \hbox{$>$}}{\lower 0.511ex \hbox{$\sim$}}}}
\def\simle{
    \mathrel{\rlap{\raise 0.511ex
        \hbox{$<$}}{\lower 0.511ex \hbox{$\sim$}}}}

\def\PL #1 #2 #3 {Phys. Lett.~{\bf#1} (#2) #3}
\def\NP #1 #2 #3 {Nucl. Phys.~{\bf#1} (#2) #3}
\def\NPP #1 #2 #3 {Nucl. Phys.~{\bf B} (Proc.~Suppl.)~{\bf#1} (#2) #3}
\def\ZP #1 #2 #3 {Z.~Phys.~{\bf#1} (#2) #3}
\def\PR #1 #2 #3 {Phys. Rev.~{\bf#1} (#2) #3}
\def\PP #1 #2 #3 {Phys. Rep.~{\bf#1} (#2) #3}
\def\PRL #1 #2 #3 {Phys. Rev.~Lett.~{\bf#1} (#2) #3}
\def\PTP #1 #2 #3 {Prog. Theor.~Phys.~{\bf#1} (#2) #3}
\def\MPL #1 #2 #3 {Mod. Phys.~Lett.~{\bf#1} (#2) #3}
\def\IJM #1 #2 #3 {Int. J.~Mod.~Phys.~{\bf#1} (#2) #3}
\def\EPJ #1 #2 #3 {Eur.~Phys.~J.~{\bf#1} (#2) #3}

\begin{document}

\draft

\preprint{RBRC-175}

\title{A lattice study of the nucleon excited states
with domain wall fermions}

\author{Shoichi~Sasaki${}^{\;a,b}$, 
Tom Blum${}^{\;a}$ and Shigemi Ohta${}^{\;c,a}$}

\address{
\vspace{0.5cm}
${}^{a)}$RIKEN-BNL Research Center, Brookhaven
National Laboratory, Upton, NY 11973-5000, USA}

\address{
\vspace{0.5cm}
${}^{b)}$Department of Physics,
University of Tokyo, Hongo 7-3-1, Bunkyo-ku, Tokyo 113-0033, Japan}

\address{
\vspace{0.5cm}
${}^{c)}$Institute for Particle and Nuclear Studies, KEK, Tsukuba, Ibaraki
305-0801,
Japan
\vspace{0.5cm}}

\date{
December 11, 2001}
\maketitle
\baselineskip 24pt
\begin{abstract}
\baselineskip 18pt
\indent
We present results of our numerical calculation of the mass spectrum for
isospin one-half and spin one-half non-strange baryons, \ie the
ground and excited
states of the nucleon, in quenched lattice QCD. We use a new lattice
discretization scheme for fermions,  domain wall fermions, which
possess almost exact chiral symmetry at non-zero
lattice spacing.  
We make a systematic investigation of the
negative-parity $N^*$ spectrum by using two distinct interpolating
operators at $\beta=6/g^2=6.0$ on a $16^3 \times  32  \times 16$ lattice.
The mass estimates extracted from the two operators are consistent with
each other. 
The observed large mass splitting  between this state, $N^*(1535)$,
and the positive-parity ground state, the nucleon $N(939)$, is well
reproduced by our calculations.  We have also calculated the mass of the first
positive-parity excited state and found that it is heavier
than the negative-parity excited state for the quark masses studied.

\end{abstract}

\vspace{0.5cm}
\maketitle
\newpage
\section{Introduction}

\indent\indent
An important challenge in lattice calculations
is to reproduce the hadron mass spectrum from first principles
in quantum chromodynamics (QCD). The latest lattice QCD calculations of 
the light-hadron mass spectrum in the quenched approximation agree
with experimental values within about 
5\%~\cite{{CP-PACS1},{Kim},{MILC}}.  
However, this success is mainly restricted to ground states. 
Indeed, results are scarcely available 
for the excited-state mass spectrum.

Another essential shortcoming of these calculations which use Wilson or
Kogut-Susskind fermions is the absence of chiral 
symmetry at finite lattice spacing, in accord with the 
Nielsen-Ninomiya no-go theorem~\cite{Nielsen}.
At non-zero lattice spacing
Wilson fermions explicitly break the full chiral symmetry of
the continuum down to the vector sub-group, so only
flavor symmetry is preserved. 
On the other hand, Kogut-Susskind fermions have only a single exact $U(1)$
axial symmetry,
and flavor symmetry is completely broken~\cite{Golterman}. Of course, it
is expected that in the continuum limit, which is difficult to achieve in
practice, both actions recover the full
chiral symmetry. 

Several years ago Kaplan constructed a new 
type of lattice fermion~\cite{Kaplan} known as domain wall fermions,
which were further developed
by Shamir~\cite{Shamir1,Shamir2} and also by Narayanan 
and Neuberger~\cite{Narayanan}. Especially, the former
reformulated it for lattice QCD simulations.
The key feature of domain wall fermions is that they utilize an extra fifth
dimension 
to circumvent the Nielsen-Ninomiya no-go theorem and maintain chiral symmetry
at non-zero lattice spacing. In practical simulations the extra dimension
is finite, so the chiral symmetry is not exact. The symmetry breaking is
very soft, however, since it is highly suppressed with the number of sites
in the extra dimension, $L_s$. 
In other words, $L_s$ gives us a way to control the violation of 
chiral symmetry. Domain wall fermions also possess exact flavor
symmetry for any value of $L_s$.

Quenched lattice QCD calculations with domain wall fermions have shown that
good chiral properties are obtained for moderate sizes of the
fifth dimension, $L_s\sim 10-16$, 
if the lattice spacing is small enough ($a \simle 0.1$
fm)~\cite{Blum1,DWF}. Recent studies by the RIKEN-BNL-Columbia-KEK 
collaboration~\cite{RBC1} and the CP-PACS collaboration~\cite{CP-PACS2} have
quantified in detail the explicit chiral symmetry breaking effects due
to finite $L_s$. For low energy QCD, the results can be simply summarized: 
there is a unique additive quark mass, $m_{\rm res}$, appearing in 
the low energy QCD lagrangian which
arises from the finite size of the extra dimension~\cite{RBC1}. This additive
quark mass has been measured quite accurately, and for $L_s=16$ and quenched
$\beta=6.0$, $m_{\rm res}\approx$ 3\% of the 
strange quark mass\cite{RBC1,CP-PACS2}. 
In this paper, we apply domain wall fermions to
baryon excited states, especially the spin one-half and 
isospin one-half negative-parity nucleon, $N^*(1535)$, as a further test of 
domain wall fermions in the baryon sector.
For masses which are $O(1)$ GeV in
the chiral limit, we do not expect $m_{\rm res}$ to have a significant effect.

We are interested in a long-standing puzzle in 
the excited state spectrum of the nucleon.
The first question addressed in this paper is whether 
the mass difference between the nucleon $N(939)$ and the negative-parity 
nucleon $N^*(1535)$ is well reproduced in lattice QCD.
The spin one-half $N^*$ state can be considered the 
{\it parity partner} of the nucleon. Of particular 
interest is the large mass splitting between $N$ and $N^*$.
From the viewpoint of parity partners, these two states would
be degenerate if the relevant 2-flavor chiral
symmetry were exact and preserved by the vacuum~\cite{Pagels}.
Of course, there is no proof that this large mass splitting 
comes directly from the spontaneous 
breaking of chiral symmetry. 
However, at least spontaneous chiral symmetry 
breaking is responsible for 
the absence of such parity doubling since the explicit 
breaking is quite small in the case of two flavors.
In this sense, regardless of a model or a theory, 
{\it chiral symmetry and its spontaneous breaking} 
are important for reproducing 
precisely the mass splitting between parity-partner hadrons.
In this paper we show that domain wall fermions accurately reproduce 
the large observed mass splitting
(some of our results have been reported earlier~\cite{Sasaki,Ohta}).

Conventional lattice fermion schemes have had
difficulty in this challenge. 
An early calculation~\cite{Gusken} using Wilson fermions 
as well as a more recent one~\cite{Gupta}, both just managed 
to extract a mass splitting between the parity partners
with large uncertainties.\footnote{In a recent preprint \cite{Richards3}
which appeared after this work was complete
it is clear that the chiral symmetry of their
fermions was significantly improved  
by simulating closer to the continuum limit than previous studies which used
Wilson fermions and by adding the clover term.}

The first sophisticated calculation\cite{FXLee1}
using an improved Wilson fermion action and relatively heavy quarks 
resolved a definite mass splitting between the parity partners which, however,
was about a factor of two smaller than the observed splitting. 
Recent results using improved Wilson fermions have confirmed 
the large mass splitting over a wide range of quark masses that we have 
found with domain wall fermions\cite{FXLee2,Richards2}.
We note that the leading lattice spacing errors that are removed 
from these improved calculations break chiral symmetry.  
Although Kogut-Susskind fermions have
a remnant $U(1)$ axial symmetry, they cannot be used practically for the
spin one-half $N^*$ mass calculation due to flavor mixing.
The reason is that Kogut-Susskind fermions have 
only discrete flavor symmetries belonging to a subgroup of 
$SU(4)$~\cite{Golterman} which contains 
three irreducible representations, {\bf8}, {\bf 8$'$} and {\bf 16} for 
baryon operators. 
Two appropriate representations {\bf 8} and {\bf 16}, 
to which $N^*$(1535) belongs, also contain the 
negative-parity $\Lambda$ states
$\Lambda$(1405) and $\Lambda$(1520) and the spin three-half 
negative-parity nucleon state $N^*$(1520)~\cite{Golterman}.  
Thus, the study of the spin one-half $N^*$ spectrum with Kogut-Susskind 
fermions always faces inevitable contamination from lower mass states.

It is also interesting to note that a non-relativistic quark model
with the so-called color magnetic interaction~\cite{Isgur}
and the MIT bag model~\cite{Bag}, both of which explicitly break chiral 
symmetry, have some difficulty reproducing the large mass splitting
between $N(939)$ and $N^*(1535)$ without adoption of less realistic
model parameters. The non-relativistic quark model is based on a harmonic
oscillator description of the orbital motion of constituent quarks.  
As remarked in the appendix of the original paper by Isgur and Karl~\cite{Isgur},
the plausible value of its oscillator quantum should be roughly 250 MeV 
to reproduce the observed charge radius and magnetic moment of the nucleon.
Since this model regards $N^*(1535)$ as a state with one quantum 
excitation in orbital motion, it indicates that the corresponding 
$N^*$ state lies at most a few hundred MeV above the ground state.
Even worse, there is the serious problem of the wrong ordering 
between $N^*(1535)$ and the positive-parity excited nucleon 
$N'(1440)$ because the corresponding 
$N'$ state should be assigned two oscillator 
quanta in this model~\cite{Isgur}.
This wrong ordering problem does not seem to be easily 
alleviated~\cite{Glozman}. It is easy to see that the MIT bag model 
faces essentially the same problem, as the single quark states 
in the model alternate in parity with roughly even spacings~\cite{Bag}.

The question arises how does this ordering of $N^*$ and $N'$ appear in 
lattice QCD calculations? 
Until \cite{Sasaki} this question could not be answered for lack of 
systematic results and investigations. 
A calculation of the mass of the positive-parity excited 
nucleon is, of course, much more difficult than the nucleon ground-state. 
Attempts have been made to evaluate the $N'$ mass 
from a two state fit to the nucleon correlation 
function~\cite{NPRIME}. However, large statistics are required compared to 
a single exponential fit. Also it is difficult to control the systematic errors.
Leinweber used the QCD sum rule (QCDSR) continuum model and
quenched lattice data to estimate the threshold of contributions
from excited states~\cite{Leinweber}.  
However, the scaling of his result has been questioned 
by Allton and Capitani~\cite{Allton}.
In this paper we take an alternative approach,
using the continuum-like behavior of the 
domain wall fermion operator~\cite{Sasaki}
to obtain the mass of the positive-parity excited nucleon. 
Our results have been confirmed in \cite{FXLee2} where
a similar approach is employed in the context of improved Wilson fermions.

In conventional lattice QCD calculations
an interpolating operator which is strongly associated with the 
non-relativistic limit is used to extract the 
mass of the nucleon ground state.
A second unconventional operator, which does not have a 
non-relativistic limit, is discarded since it is expected 
to couple weakly to the ground state.
Indeed, using this operator, no one has succeeded in evaluating
the mass of the nucleon ground state in lattice QCD calculations 
with Wilson fermions~\cite{Bowler} because of its
small coupling to the ground state~\cite{Leinweber}.
The expectation of an approximately zero
overlap on the ground state provides the possibility that the 
use of the second operator in lattice calculations 
directly yields the mass of the $N'$ state, at least 
in the relatively heavy (valence) quark 
mass region. This prospect, however, is built on the assumption that
the lattice defined operators inherit the features 
of the continuum ones. In the case of the Wilson fermions, the Wilson term, 
which explicitly breaks chiral symmetry, induces mixing between the
conventional and unconventional operators~\cite{{Richards1},{Leinweber}}. 
Thus, the desired feature of the unconventional operator in the continuum 
is diminished in lattice calculations with Wilson fermions.
On the other hand, this type of mixing between three quark operators 
is absent at one loop in perturbation theory using domain wall
fermions with large $L_s$~\cite{Aoki}; thus we expect this mixing to be
suppressed in domain wall fermion lattice calculations.
Indeed, we find that the second operator
has negligible overlap with the ground state
and furthermore provides a reliable signal of the excited state 
as the asymptotic state in the heavy quark mass region.

The organization of our paper is as follows. In Sec. II, we briefly 
review the basic formulae and notation regarding domain wall fermions.
In Sec. III we investigate the properties of the two-point 
correlation function for the nucleon and its parity partner.
Sec. IV gives the details of our Monte Carlo simulations and the 
results for the parity partner of the nucleon ($N^*$) and the
first positive-parity excited nucleon ($N'$). Then, we compare our results to
the experimental values. Finally, we present our conclusions in Sec. V.

\section{Domain wall fermions}

\indent\indent 
In this section we closely follow the development of domain wall fermions
by Shamir~\cite{Shamir1,Shamir2}.
The domain wall fermion action is essentially regarded as
a five-dimensional extension of the Wilson fermion action:
%
%
\beqn
S_{\rm DWF}=-\sum_{x,x'}\sum_{s,s'}{\bar 
\Psi}(x,s)[\delta_{s,s'}D^{\parallel}_{x,x'}+\delta_{x,x'}D^{\perp}_{s.s'}]
\Psi(x',s')\;,
\eeqn
where $x,x'$ are four-dimensional Euclidean space-time coordinates and
$s,s'$ denote coordinates in the extra dimension labeled from 0 to $L_{s}-1$
(to take advantage of existing high-speed computer code, our domain
wall fermion Dirac operator is
the Hermitian conjugate of the one in~\cite{Shamir2},
hence our notation is the same as in~\cite{RBC1}).
Here, $D^{\parallel}_{x,x'}$ corresponds to the four-dimensional
Wilson-Dirac operator
with a mass term (domain wall height) $M_{5}$.
%
%
\beqn
D^{\parallel}_{x,x'}=\frac{1}{2}\sum_{\mu}\left\{(1-\gamma_{\mu})U_{\mu}(x)
\delta_{x+{\hat \mu},x'}
+(1+\gamma_{\mu})U^{\dag}_{\mu}(x')\delta_{x-{\hat \mu},x'}\right\}
+(M_{5}-4)\delta_{x,x'}\;,
\eeqn
where the Wilson term and mass term
have opposite relative sign compared to the 
conventional one. 
$D^{\perp}_{s,s'}$ is the five-dimensional analog of the four dimensional
Wilson hopping term with $\gamma_\mu$ replaced by $\gamma_5$ and $U_5(x,s)=1$.
%
%
\barr
D^{\perp}_{s,s'}&=&\frac{1}{2}\left\{(1-\gamma_{5})\delta_{s+1,s'}
+(1+\gamma_{5})\delta_{s-1,s'}-2\delta_{s,s'}\right\} \nonumber \\
&&-\frac{m_{f}}{2}\left\{(1-\gamma_{5})\delta_{s,L_{s}-1}\delta_{0,s'}
+(1+\gamma_{5})\delta_{s,0}\delta_{L_{s}-1,s'}\right\}\;.
\earr
Note that the five dimensional 
fermions $\Psi(x,s)$ are coupled only to four-dimensional gauge fields.
The boundaries $s=0$ and $s=L_{s}-1$ are anti-periodic and coupled with 
a weight $m_{f}$. 

For an appropriate choice of $M_5$, this action has two chiral zero-modes
of opposite handedness, one
localized on each boundary of the fifth
dimension. To simulate low energy QCD, 
four dimensional quarks are interpolated from the chiral modes,
%
%
\barr
q(x)&=&P_{L}\Psi(x,0)+P_{R}\Psi(x,L_{s}-1)\;, \\
{\bar q}(x)&=&{\bar \Psi}(x,L_{s}-1)P_{L}+
{\bar \Psi}(x,0)P_{R}\;,
\label{eq:4d-quark}
\earr
where $P_{R,L}=(1\pm \gamma_{5})/2$. With this definition,
one can see that the $m_{f}$ terms in $D^{\perp}_{s,s'}$ 
directly yield the usual four dimensional 
explicit chiral symmetry breaking term, 
$m_{f}{\bar q}(x)q(x)$, on the four dimensional layers $s=0$ and $s=L_{s}-1$.
The above definition of the quark fields 
is the simplest one but is not unique. 

In the free theory, for the choice $0<M_{5}<2$, the domain wall fermion action
corresponds to one flavor, and the four-dimensional light quark mass
is given as 
$m_{q}=m_{f}M_{5}(2-M_{5})$ in the limit 
$L_{s}\rightarrow \infty$ and $m_{f} \rightarrow 0$. 
This situation is approximately unchanged in lattice QCD simulations 
if $M_{5}$ is simply shifted, $M_{5}\rightarrow 
M_{5}-M_{c}$~\cite{Blum1}, where a simple estimate of $M_{c}$ is 
given in terms of the critical hopping parameter for 
four-dimensional Wilson fermions~\cite{Blum2}.

In the case of
domain wall fermions, the $SU(N_{f})$ axial transformation can be 
defined vectorially~\cite{Shamir2} as
%
%
\barr
\left[\right.{\cal Q}^{a}_{A}, \Psi(x,s)\left.\right] &=&
+i\epsilon(s)\lambda^{a}
\Psi(x,s)\;, \label{eq: 5d-AxialT1}\\
\left[\right.{\cal Q}^{a}_{A}, {\bar \Psi}(x,s)\left.\right] &=& -i\epsilon(s)
{\bar \Psi}(x,s)\lambda^{a}, \label{eq: 5d-AxialT2}
\earr
where $\Psi(x,s)$ is a five-dimensional zero mode. 
Opposite axial charges are assigned to fermions in the two 
half-spaces, so $\epsilon(s)=1$ for $0 \leq s < L_{s}/2$ and $\epsilon(s)=-1$ 
for $L_{s}/2 \leq s \leq L_{s}-1$. The reason for this is clear: left and
right handed modes are globally separated in the
extra dimension.

With this axial transformation on the five dimensional fields, 
the four dimensional quark fields 
obey the familiar axial transformation:
%
%
\barr
\left[\right.{\cal Q}^{a}_{A}, q(x)\left.\right] &=& 
+i\gamma_{5}\lambda^{a}q(x)\;, \\
\left[\right.{\cal Q}^{a}_{A}, {\bar q}(x)\left.\right] &=& 
-i{\bar q}(x)\gamma_{5}\lambda^{a}\;. \label{eq: 4d-AxialT}
\earr

In general the domain wall fermion action is not invariant
under the transformations (\ref{eq: 5d-AxialT1}) and (\ref{eq: 
5d-AxialT2}) in the limit $m_{f}=0$ 
because of a non-vanishing divergence of the axial current on the
intermediate layers $s=L_{s}/2-1$ and $s=L_{s}/2$ which 
gives rise to an extra pseudo-scalar density
in the axial Ward-Takahashi identity~\cite{Shamir2}. 
However, such an anomalous term, which provides a chiral symmetry breaking
effect due to mixing of the left- and right-handed modes, 
vanishes as $L_{s}\rightarrow \infty$~\cite{Shamir2}.

Theoretically, this residual breaking effect can 
be described by an additive
quark mass $m_{\rm res}$ in the four-dimensional 
low energy effective Lagrangian for QCD~\cite{RBC1}, and recent simulations,
in which $m_{\rm res}$ was determined in several ways, appear to confirm 
this\cite{RBC1}. Simulations
also show that for sufficiently small lattice spacing, $m_{\rm res}\to 0$ as $L_s\to\infty$, and at the
very least, $m_{\rm res}$ is small and accurately known for a wide range of 
$L_s$\cite{RBC1,CP-PACS2}.
Thus, in this study we will henceforth ignore these
small effects and assume the chiral limit is $m_f=0$ instead of 
$m_f=-m_{\rm res}$. Finally, we note that the 
non-perturbative origin of these effects and
their relation to certain non-perturbative gauge field configurations
is a very interesting and active area of research~\cite{EVERYONE}.

\section{Baryon spectrum}
\subsection{Baryon two-point correlator}
\indent\indent
The mass $m_{B}$ of the low-lying baryon $B$ can be extracted from the 
two-point correlation function 
composed of the baryon interpolating operator ${\cal O}_{B}$,
which has the appropriate quantum numbers specified by the desired 
state.
Let us consider the vacuum expectation value of the time-ordered 
product of interpolating operators. The Euclidean time correlation
function
is projected out at zero spatial momentum through the sum over 
$\vec x$:
%
%
\beqn
G_{{\cal O}_{B}}(t)=\sum_{\vec{x}}\lvac T\{ {\cal O}_{B}(\vec{x},t) 
{\bar{\cal O}}_{B}(0,0) \}\rvac \;,
\eeqn
which is dominated by the contribution of the lowest mass state 
for large Euclidean time,
$G_{{\cal O}_{B}}(t)\sim\exp(-M_{B}|t|)$.
For a spin one-half baryon, of course, the
correlation function has non-trivial Dirac structure which may be 
expressed in the form
%
%
\beqn
G_{{\cal O}_{B}}(t) 
\mathrel{\mathop{\kern0pt\longrightarrow}\limits_{{\rm large}\;\;t}}
 (1+\gamma_{4})\theta(t)A_{B}e^{-M_{B}t}
+ (1-\gamma_{4})\theta(-t)A_{B}e^{+M_{B}t}\;,
\eeqn
where the positive definite $A_{B}$ is proportional to 
the square of the coupling strength 
between the interpolating operator ${\cal O}_{B}$ 
and the lowest mass state. 
The first and second terms are {\it particle} and {\it anti-particle}
contributions, respectively. It is easy to get only the particle 
contribution by taking the trace with 
the projection operator $P_{+}=(1+\gamma_{4})/2$, which we abbreviate 
as $\langle\langle{\cal O}_{B}(t){\bar {\cal O}}_{B}(0)\rangle\rangle
={\rm Tr}\{P_{+}G_{{\cal O}_{B}}(t)\}$ hereafter.

\subsection{Interpolating operators}
\indent\indent
Let us focus on the nucleon channel specified by 
the spin one-half iso-doublet non-strange baryons. 
In order to project out the desired channels, we have to construct
interpolating operators from the quark fields
with the appropriate quantum numbers ($J=1/2$ and $I=1/2$). 
However, there is considerable freedom in choosing the specific form of the
composite operators.
Indeed, there are two possible interpolating operators for the
$J^P = 1/2^+$ state, even if we restrict them
to contain no derivatives and to belong to the
$(\frac{1}{2},0)\oplus(0,\frac{1}{2})$ 
chiral multiplet under $SU(2)_{L}\otimes SU(2)_{R}$~\cite{Ioffe,Cohen}:
%
%
\barr
B^{+}_{1}(x)&=&\varepsilon_{abc}[u^{T}_{a}(x)C\gamma_{5}d_{b}(x)]u_{c}(x)\;, 
\label{eq:B1+}\\
B^{+}_{2}(x)&=&\varepsilon_{abc}[u^{T}_{a}(x)Cd_{b}(x)]\gamma_{5}u_{c}(x)\;,
\label{eq:B2+}
\earr
where $abc$ and $ud$ have usual meanings as color and
flavor indices. $C$ is the charge conjugation matrix and
the superscript $T$ denotes transpose. Here, Dirac indices have been 
suppressed. In Eq.(\ref{eq:B1+}) and (\ref{eq:B2+}), the superscript 
``$+$'' refers to positive parity since these operators transform as 
${\cal P}B^{+}_{1,2}(\vec{x},t){\cal P}^{\dag}=+\gamma_{4}B^{+}_{1,2}
(-\vec{x},t)$ under parity.
To be precise, the linear combinations
$B_{1}^{+}\pm B_{2}^{+}$
belong to distinct $(\frac{1}{2},0)\oplus(0,\frac{1}{2})$ 
chiral multiplets under $SU(2)_{L}\otimes SU(2)_{R}$~\cite{Ioffe,Cohen}. 

The operator $B_{1}^+$ alone is usually used in lattice QCD calculations
to extract the nucleon ground state since 
$B_{2}^+$ couples only weakly to the ground state 
due to its vanishing in the non-relativistic limit~\cite{Leinweber}. 
In fact, nobody has succeeded in extracting the nucleon mass spectrum
from the $B_{2}^+$ operator. In our calculation, 
we also confirm that the ground state cannot be extracted
from
$\langle\langle B_{2}^{+}(t){\bar B}_{2}^{+}(0)\rangle\rangle$;
however, we have had some success with respect to the excited-state mass
spectrum which we discuss in Sec IV-C.

Multiplying the left hand side of the previous positive-parity operators 
by $\gamma_{5}$, we obtain the interpolating 
operators with negative parity, $J^P = 1/2^-$~\cite{FXLee1}:
%
%
\barr
B^{-}_{1}(x)&=&\gamma_{5}B^{+}_{1}(x)=\varepsilon_{abc}[u^{T}_{a}(x)C\gamma_
{5}d_{b}(x)]
\gamma_{5}u_{c}(x)\;, \\
B^{-}_{2}(x)&=&\gamma_{5}B^{+}_{2}(x)=\varepsilon_{abc}[u^{T}_{a}(x)Cd_{b}(x
)]u_{c}(x)\;,
\earr
since ${\cal P}B^{-}_{1,2}(\vec{x},t)
{\cal P}^{\dag}=-\gamma_{4}B^{-}_{1,2}(-\vec{x},t)$.
The important point to notice is the relation between the 
correlation functions of opposite parities,
%
%
\beqn
G_{B^{+}}(t) = -\gamma_{5}G_{B^{-}}(t)\gamma_{5}\;,
\label{eq:posneg}
\eeqn
since $B^{-}_{1,2}=\gamma_{5}B^{+}_{1,2}$.
Eq.(\ref{eq:posneg}) means that the two-point
correlation function of the spin one-half baryon can couple 
to both positive- and negative-parity states. 
Thus, the general form of the two point function is ~\cite{Fucito}
%
%
\barr
G_{B^{+}}(t)&=& (1+\gamma_{4})\theta(t)A_{B^{+}}e^{-M_{B^+}t} + 
                (1-\gamma_{4})\theta(-t)A_{B^{+}}e^{+M_{B^+}t}\nonumber\\
        &-& (1+\gamma_{4})\theta(-t)A_{B^{-}}e^{+M_{B^-}t} - 
                (1-\gamma_{4})\theta(t)A_{B^{-}}e^{-M_{B^-}t}\;.
\label{eq:BaryonFun}
\earr
Note that
the backward propagating contributions correspond to the anti-particles
of the forward propagating states with opposite parity. The desired state
is obtained by choosing the appropriate projection operator, $1\pm\gamma_4$,
and direction of propagation.

Next, consider a lattice with finite extent $T$ in the time direction and 
(anti-)periodic boundary conditions.
Eq.~\ref{eq:BaryonFun} is replaced by 
%
%
\barr
G_{B^{+}}(t)&=& (1+\gamma_{4})A_{B^{+}}e^{-M_{B^+}t} \pm 
                (1-\gamma_{4})A_{B^{+}}e^{-M_{B^+}(T-t)}\nonumber\\
        &\mp& (1+\gamma_{4})A_{B^{-}}e^{-M_{B^-}(T-t)} - 
                 (1-\gamma_{4})A_{B^{-}}e^{-M_{B^-}t}\;,
\earr
where the (lower) upper sign stands for (anti-)periodic boundary 
condition on $0 < t < T$.
The anti-particle of the opposite-parity
state can propagate through the time direction boundaries,
so one faces unwanted contamination 
from the opposite-parity state in extracting
the mass of the desired state.
This contamination is not unavoidable 
if a double exponential fit is used. 
However, such fits require 
very high statistics.
This is not a serious issue in the measurement 
of the nucleon ground state since the 
contamination from the negative-parity nucleon is 
expected to be negligible due to the large mass splitting
$M_{B^{-}}-M_{B^{+}}$. Of course, this same splitting does affect
the extraction of $M_{B^-}$.
The problem is resolved by choosing 
appropriate boundary conditions
in the time direction to prevent the wrap-around effect, or by increasing
the time extent $T$ sufficiently and placing the 
interpolating operators far from the boundary.
In this study we take the former approach.
It is common to employ Dirichlet boundary conditions where link valuables
in the time direction
at $t=0$ and $t=T-1$ are set to zero when calculating quark propagators (for
example, see \cite{FXLee1}).
However, we use a linear combination of two quark 
propagators with periodic and anti-periodic boundary conditions in the time 
direction to produce forward propagating states.
Although our approach requires two times as many fermion matrix inversions to 
calculate one quark propagator, it does not suffer from (unknown) reflection 
effects at the time boundaries induced by Dirichlet boundary conditions. 

\subsection{Chiral symmetry and parity doubling}
\indent\indent
At the end of this section, let us briefly review how
unbroken chiral symmetry imposes parity doubling in
the hadron spectrum~\cite{{Pagels},{Cohen}}. 
We will generalize the following argument in Appendix A.
For the sake of simplicity, we consider a particular 
transformation of the $SU(2)$ chiral symmetry,
$[{\cal Q}_{A}, u]=+i\gamma_{5}u$ and $[{\cal Q}_{A}, d]=-i\gamma_{5}d$.
Then, one can easily find that the two-point correlators
$B_{1}^{\pm}$ and $B_{2}^{\pm}$ transform as
%
%
\beqn
\left[ {\cal Q}_{A}, B^{\pm}_{1,2}(x){\bar 
B}^{\pm}_{1,2}(0)\right]= 
i \left\{ \gamma_{5}, B^{\pm}_{1,2}(x){\bar B}^{\pm}_{1,2}(0)\right\}
\label{eq:trans}
\eeqn
in the chiral limit.
Now suppose that the vacuum possesses 
chiral symmetry: ${\cal Q}_{A}\rvac = 0$. 
According to Eq.(\ref{eq:trans}),
the two-point correlation functions
anti-commute with $\gamma_5$,
%
%
\beqn
\left\{ \gamma_{5}, G_{B^{\pm}}(t)\right\} = 0\;.
\eeqn
Immediately, with the help of Eq.(\ref{eq:posneg}), one finds 
%
%
\beqn
G_{B^{+}}(t)=G_{B^{-}}(t) \;,
\eeqn
which means that parity doubling arises 
in the nucleon channel due to chiral symmetry~\cite{Cohen}. 
Of course, in the real world chiral symmetry is spontaneously 
broken, ${\cal Q}_{A}\rvac \ne 0$, so that such 
parity doubling does not occur in the actual spectrum~\cite{Pagels}. 
In this sense, the spontaneous breaking of chiral symmetry is 
responsible for the absence of parity doubling. 
Thus, it seems important to properly handle chiral symmetry
and its spontaneous breaking in order to calculate 
precisely the mass splitting between the nucleon and its 
parity partner.
Finally, it is important to note that above argument 
ignores possible consequences to the 't Hooft anomaly 
condition~\cite{tHooft,Norman}.

\section{Numerical results}

\subsection{Computational details}
\indent\indent
We generate quenched QCD configurations on a $16^3 \times 32$ 
lattice with the standard single-plaquette Wilson action 
at $\beta=6/{g^2}=6.0$. The quark propagator is calculated 
using domain wall fermions with a fifth dimension of $L_{s}$=16 
sites and a domain wall height $M_{5}$=1.8. 
Additional details of the simulation can be found in \cite{RBC1,RBC2}.
Quenched $\beta=6.0$ corresponds to a lattice cut-off of
$a^{-1}\approx 1.9$ GeV from $aM_{\rho}$=0.400(8) in
the chiral limit and spatial size $La \approx 1.7$ fm~\cite{RBC1,RBC2}.

We work in Coulomb gauge 
and calculate quark propagators using wall sources
and local sinks.
We expect that the use of wall sources provides better 
overlap with the desired states. 
To extract the state with desired parity in the spin one-half baryon 
spectrum, 
we construct forward (backward) propagating quarks by taking the
appropriate linear combination of propagators with periodic and
anti-periodic boundary conditions 
in the time direction, as mentioned before. 
This procedure eliminates the
backward (forward) propagating opposite parity state which wraps around
the time boundary.

In addition, we use two sources for quark propagators on each configuration
($t_{\rm src}$ = 5 and $t^\prime_{\rm src}$ = 27) to increase statistics.
After the appropriate parity projections, the correlation functions
of the corresponding nucleon state can be folded together (averaged
as a function of distance from the respective source).
Here, time-slices are labeled from 0 to 31. 
To be clear, we give the form of the nucleon two-point function with 
arbitrary source location $t_{\rm src}$
after eliminating the unwanted contributions across time boundaries:
%
%
\beqn
G_{B^{+}}(t-t_{\rm src}) = \left\{
\begin{array}{ll}
(1+\gamma_{4})A_{B^{+}}e^{-M_{B^{+}}(t-t_{\rm src})}
-(1-\gamma_{4})A_{B^{-}}e^{-M_{B^{-}}(t-t_{\rm src})} & (T> t > t_{\rm 
src})\;, \\
(1-\gamma_{4})A_{B^{+}}e^{-M_{B^{+}}(t_{\rm src}-t)}
-(1+\gamma_{4})A_{B^{-}}e^{-M_{B^{-}}(t_{\rm src}-t)}
& (t_{\rm 
src}>t>0)\;, 
\end{array}
\right.
\eeqn
which is constructed with the positive-parity interpolating operator
(either $B_{1}^{+}$ or $B_{2}^{+}$).
In the time range $T> t > t_{\rm src}$, positive- and negative-parity nucleon 
states are propagating only forward in time. This means that there are only 
particle contributions in this region. Thus, $P_{\pm}=(1\pm 
\gamma_{4})/2$ 
project out positive- and negative-parity states, 
respectively. Alternatively, for $t_{\rm 
src}>t>0$ only the anti-particles contribute to the correlation function.
In this case, $P_{\pm}$
act in the opposite way to the former case.
In either case, all we have to do is calculate either $G_{B^+}$ or $G_{B^-}$ 
to extract the masses for both positive- and negative-parity states
since we already know ${\rm Tr}\{P_{\pm}G_{B^+}(t)\}=
-{\rm Tr}\{P_{\mp}G_{B^-}(t)\}$ with the help of Eq.\ref{eq:posneg}.
In fact, we verified this relation on
each configuration and then used $G_{B_{1}^+}$ and $G_{B_{2}^-}$ 
to extract masses.  

In our analysis, we use correlation functions in the ranges $t > 5$ and
$t<27$, for the sources $t_{\rm src}=5$ and $t^\prime_{\rm src}=27$,
respectively.
After folding the propagators together as described above, we use a
single offset from the source, $t$, defined in the
range $0 < t < 27$

We use 405 independent gauge configurations for the lightest 
two quark masses, $m_{f}=0.02$ and $0.03$, 305 configurations for 
the intermediate ones, $m_{f}=0.04$ and $0.05$, and 105 configurations
for the heavier ones, $m_{f}=0.075-0.125$. These bare quark masses 
correspond to mass ratios $M_{\pi}/M_{\rho} 
\approx 0.59 - 0.90$ as shown in Table~\ref{tab:massratios}.
In this calculation, $SU(2)$-isospin symmetry
is enforced by equating $m_{f}=m_{f}^{(u)}=m_{f}^{(d)}$,
so that the flavor index will not be explicitly displayed hereafter.
All calculations were done on the 600 Gflops QCDSP machine at 
the RIKEN BNL Research Center.

\subsection{Parity partner of nucleon: $N^*$}
\indent\indent
We first calculate the effective 
masses $M_{\rm eff}$ to find appropriate time ranges for fitting. 
The effective mass is defined by
%
%
\beqn
M_{\rm eff}(t)=\ln\{C(t)/C(t+1)\}\;,
\eeqn
where $C(t)$ stands for 
$\langle\langle B_{1,2}^{\pm}(t){\bar B}_{1,2}^{\pm}(0)\rangle\rangle$.
We look for a plateau, or time independent region, in this quantity to
extract the ground state mass.
For example, Figures 1(a)-1(c) show effective masses for 
the nucleon ($B_{1}^{+}$) and its parity partner ($B_{1,2}^{-}$)
at $m_{f}=0.03$, 0.05, and 0.10.
In Figure 1, 
the effective mass plot shows a clear plateau for the $N$,  
and one that is not as good for the heavier $N^*$. 
Statistical uncertainties in Figure 1 are estimated by a
single elimination jack-knife method.
The effective masses for the $N^*$ from both
$\langle\langle B_{1}^{-}(t){\bar B}_{1}^{-}(0)\rangle\rangle$ and
$\langle\langle B_{2}^{-}(t){\bar B}_{2}^{-}(0)\rangle\rangle$
agree well with each other for all three 
quark masses, except for $m_f=0.03$ and $t\simle 6$ where there
is a small difference outside of statistical errors.

In Figure 1 (a), the effective mass
for the $N^*$ becomes so noisy after $t=9$ that the value of the 
two-point correlator is consistent with zero within large errors, so we have 
left these points off of the plot. In addition, the effective 
mass shows a steep rise 
with increasing uncertainty as $t$ increases. 
This rise in the $N^*$ effective mass 
weakens at relatively heavy quark masses where 
the signal becomes stable over 12 time-slices.

Next we present mass estimates of the $N$ and $N^*$ obtained from
covariant single exponential fits to the corresponding correlators. 
We fit each correlator from some 
minimum time-slice, $t_{\rm min}$, to an appropriate maximum time-slice 
($t_{\rm max}=20$ for the $N$ and $t_{\rm max}=10-15$ for the 
$N^*$).
$t_{\rm max}$ is roughly fixed with reference 
to the effective mass calculation. 
To keep fitting ranges as wide as possible,
$t_{\rm min}$ is reduced from $t_{\rm max}-2$ until
$\chi^2 / N_{\rm DF} > 1.5$ where $N_{\rm DF}$
denotes the degrees of freedom in the fit. Fitting details
are given in Tables~\ref{tab:fit1}-\ref{tab:fit3}.
All of our fits have confidence-level larger than 0.2 and estimates 
from the weighted average of the effective mass agree with 
the fitted masses within errors. 
A summary of our $N$ and $N^*$ masses is given 
in Tables~\ref{tab:fit1}-\ref{tab:fit3}.

We can roughly estimate the systematic error coming from 
the choice of fitting range by varying $t_{\rm min}$.
Let us determine the difference between our final fits and fits 
where the smallest time slice is not included. In the case of 
$B_{1}^{+}$ (the nucleon), resulting errors are much smaller than the 
statistical errors of the final fits. On the other hand, for both 
$B_{1}^{-}$ and $B_{2}^{-}$, the estimated systematic errors are
comparable to the statistical errors, while for $B_{1}^{-}$ at
$m_{f}=0.04$ and 0.05 the systematic error is a factor of two larger than
the statistical one. This uncertainty is directly related to the rise
of the $N^*$ effective mass, or poor plateau, in those cases
where the correlation function was not statistically well resolved at
larger times.
This, in turn, has forced us to extract masses from the smaller $t$ region which may
be contaminated by excited states.

In Figure 2
we show the low-lying nucleon spectrum as a function of the quark 
mass, $m_{f}$. The nucleon mass is extracted from $B_{1}^{+}$.
We omit the point at $m_{f}=0.02$ for the operator $B_{2}^-$ since 
a good plateau in the effective mass plot is absent.
The $N^*$ mass estimates from
$\langle\langle B_{1}^{-}(t){\bar B}_{1}^{-}(0)\rangle\rangle$ and
$\langle\langle B_{2}^{-}(t){\bar B}_{2}^{-}(0)\rangle\rangle$
agree with each other within errors in the whole quark mass 
range, as expected from their common quantum numbers~\cite{Sasaki}.
Note that this result disagrees
with that obtained in \cite{FXLee1}:
we find no discrepancy between masses extracted from 
$\langle\langle B_{1}^{-}(t){\bar B}_{1}^{-}(0)\rangle\rangle$ and
$\langle\langle B_{2}^{-}(t){\bar B}_{2}^{-}(0)\rangle\rangle$.
However, recent results in \cite{FXLee2} and
\cite{Richards2} are in good agreement with ours.
We also obtain the same mass for the $N^*$ from a mixed 
correlator $\langle\langle B_{1}^{-}(t){\bar B}_{2}^{-}(0)+B_{2}^{-}(t){\bar 
B}_{1}^{-}(0)\rangle\rangle$. 

The most remarkable feature in Figure 2, which was first 
reported in \cite{Sasaki}, is that 
the $N$-$N^*$ mass splitting 
is observed over the whole range of quark mass values and
grows as the quark mass is decreased.
To illustrate this point clearly,
we compare two mass ratios in Figure 3, one from the baryon parity partners
$M_{N^*}/M_N$ and the other from pseudo-scalar and vector mesons 
$M_\pi/M_\rho$.
Experimental points~\cite{PDG} are marked with stars, 
corresponding to non-strange (left) and strange (right) sectors. In 
the strange sector we use $\Sigma^{+}$ and $\Sigma(1750)$ as baryon 
parity partners and $K$ and $K^*$ for the mesons. 
The baryon mass ratio clearly grows with decreasing meson mass ratio, 
toward the experimental values~\cite{Sasaki}.
We did not include the charm sector
($M_{D}/M_{D^{*}}\simeq 0.93$) since the 
parity partner of $\Sigma_{c}^{++}(2455)$ is not measured experimentally.
On the other hand, from our results we estimate the mass of this state 
to be roughly 2.7 GeV (we have used a simple linear in the quark mass ansatz 
to extract the mass from our degenerate quark data).

Finally, we evaluate the $N$ and $N^*$ masses in the chiral limit.
Taking a simple linear extrapolation in the four lightest quark masses 
for $B_{1}^{+}$ and $B_{1}^{-}$, we find $M_N$=0.57(1) and $M_{N^*}$=0.85(5)
in lattice units. 
Setting the scale from the calculated \(\rho\)-meson mass \cite{RBC1}, we
obtain $M_{N}\approx 1.1$ GeV and $M_{N^*}\approx 1.6$ GeV in the
chiral limit.  If we use the scale set by the
calculated nucleon mass, we obtain \(M_{N^*} \approx 1.5\) GeV.
Either way the $N^*$ mass is in good agreement with the experimental
values within about 5-10 \%.
The above errors do not include systematic uncertainties due to finite volume, 
non-zero lattice spacing, and quenching effects. 
Studies of such systematic errors will be addressed in future calculations.

Recently it was suggested~\cite{Richards3}
that the $N^*$ propagator in quenched QCD may exhibit
non-analytic chiral behavior associated with the anomalous 
contribution of the ``$\eta'$-$N$'' intermediate state.\footnote{
Recall that in the quenched approximation the $\eta^\prime$ is also
a pseudo-Goldstone boson. We stress that the quenched ``$\eta'$-$N$'' 
intermediate state would give a {\it negative metric}
contribution corresponding to a ``quenched chiral loop''
artifact \cite{Bardeen}.
Needless to say, this spurious intermediate state, which is a
source of unitarity violation, is not associated with the physical
decay process \cite{Bardeen}.
}
This suggestion is inspired by recent articles \cite{Bardeen,MILC2}.
However, non-analytic effects which arise in the
vicinity of the massless quark limit are hard to detect in 
our data since our calculation is not very close to this limit. 
In addition, such subtle effects could easily 
be mimicked by finite volume
effects. The volume used in our study is still rather small.
Furthermore, the statistical errors in our study are 
large enough, especially for the $N^*$, that complicated 
fits beyond a naive linear one would not yield meaningful 
parameters or $\chi^2$.

\subsection{Unconventional nucleon operator}
\indent\indent
As mentioned earlier, the unconventional operator $B_{2}^+$ vanishes in
the non-relativistic limit,
so one may expect that it couples with negligible weight to the
nucleon ground state near this limit.
Indeed, no one has succeeded in extracting
the ground state mass signal from 
$\langle\langle B_{2}^{+} (t) {\bar B}_{2}^{+} (0) \rangle\rangle$
in lattice QCD~\cite{{Bowler},{Leinweber}}. Then, we
expect that an approximately zero overlap 
on the ground state provides the possibility of
direct access to the excited state using the $B_{2}^+$ operator.
This prospect should hold
as long as the lattice operator 
has the same symmetries as the continuum one.

In the massless quark limit, the combinations $B_{1}^{+}\pm B_{2}^{+}$
do not mix due to different chiral structure. 
In perturbation theory, each is multiplicatively 
renormalized with the same renormalization factor so that
$B_{1}^{+}$ and $B_{2}^{+}$ also do not mix.
However, conventional lattice fermions give rise 
to mixing through explicit chiral symmetry breaking~\cite{Richards1}.
Thus, $B_{2}^{+}$ will couple to the ground state through unwanted 
mixing with $B_{1}^{+}$. On the other hand, the explicit breaking of
chiral symmetry in domain wall fermions is highly suppressed, so there 
is hope that we may extract the positive-parity excited state of the nucleon 
from $B^+_2$ as the asymptotic state.

Let us first compare the effective mass plots of $\langle \langle B_{1}^{+}
(t) {\bar B}_{1}^{+} (0) \rangle \rangle$ and $\langle \langle
B_{2}^{+} (t) {\bar B}_{2}^{+} (0) \rangle \rangle$ correlators (see Figures
4a-4c).  The correlator $\langle \langle B_{2}^{+} (t) {\bar B}_{2}^{+} (0)
\rangle \rangle$ is considerably noisier so that only
time-slices near the source are useful.  Nevertheless, the effective mass from
$\langle \langle B_{2}^{+} (t) {\bar B}_{2}^{+} (0) \rangle \rangle$ yields
a plateau albeit with large statistical errors. The plateau becomes more
satisfactory for heavier quark mass.
These plateaus are obviously different from those
extracted from the $B_{1}^{+}$ correlator.  When we apply the single
exponential fit
to the two-point correlation function composed of $B_{2}^+$, we
obtain a mass that is quite large compared to the mass extracted from
$\langle\langle B_{1}^{+} (t) {\bar B}_{1}^{+} (0) \rangle\rangle$.
In \cite{FXLee2} similar results from the $B_{2}^{+}$ correlator 
have been found.

As mentioned earlier, an explanation of the above result \cite{Sasaki} is that
$B_{2}^+$ has negligible overlap with the nucleon ground state since it does
not have a non-relativistic limit and thus provides a direct signal for the
positive-parity excited state of the nucleon.  Of course, this explanation is
valid only in the heavy valence-quark mass limit.

We investigate further the possibility that $\lvac B_{2}^+ |N \rangle \approx
0$ through the calculation of  the mixed correlation function $\langle
\langle B_{1}^{+} (t) {\bar B}_{2}^{+} (0) + B_{2}^{+} (t) {\bar B}_{1}^{+} (0)
\rangle \rangle$.  If it is true that the overlap with the
nucleon $\lvac B_{2}^+ |N \rangle$ becomes small with increasing
valence-quark mass, mass estimates from the mixed correlator $\langle\langle
B_{1}^{+} (t) {\bar B}_{2}^{+} (0) + B_{2}^{+} (t) {\bar B}_{1}^{+} (0) \rangle
\rangle$ should be consistent with the nucleon for lighter quark mass and the
positive-parity excited state for heavier quark mass.  Such behavior is evident
in Figure 5. We stress that the single particle 
fit to the mixed correlation function does not in general yield the mass
of an asymptotic state, only in the limits discussed above, which
is clear from the figure. We also note that for improved Wilson
fermions the mixed correlation function yields the ground state
mass for all quark masses, even heavy ones~\cite{FXLee2}. Thus
it appears in that case that there is still significant mixing of
$B_{1}^{+}$ and $B_{2}^{+}$.  

As a result, it is possible to identify $B_{2}^{+}$
with the positive-parity excited nucleon ($N'$) for heavy quarks 
(\eg $m_{f} \simge 0.075$ in our study) \cite{Sasaki}.
Indeed, we see
$|\lvac B_{2}^+ |N\rangle / \lvac B_{2}^+ |N' \rangle |^2 \leq 10^{-3}$ from
double exponential fits for $\langle \langle B_{2}^{+} (t){\bar B}_{2}^{+} (0)
\rangle \rangle$ at $m_{f}=0.10$ and 0.125.  Of course, this feature weakens
in the lighter quark mass region (from around $m_{f}=0.05$).  However, even
for $m_{f}<0.075$, mass
estimates from $B_{2}^+$ are still considerably larger than the nucleon mass,
so we cannot rule out the possibility that $B_{2}^{+}$ provides a signal for
the positive-parity excited nucleon at even lighter quark mass values.

\subsection{Diagonalization method for 2$\times$2 matrix correlator}
\indent\indent
Let us consider the $N'$ in more detail.
The mass of the excited nucleon may be
obtained from a two state fit to the
$B_{1}^+$ correlator which, however, requires large statistics and
that neither the ground state or the first excited state 
lies close to any other state.
Attempts to extract the $N'$ mass using this method have failed 
to reproduce the observed mass~\cite{NPRIME}. 
In our fits the Hessian matrix often becomes singular,
so we cannot extract the excited state mass from double exponential fits.
Instead, we take an alternative approach proposed 
in \cite{Luscher}.
First, we define the 2$\times$2 matrix correlator ${\cal C}(t)$ using 
the two distinct baryon operators
%
%
\beqn
{\cal C}(t)=\left[
\begin{array}{cc}
c_{11}(t) & c_{12}(t) \\
c_{21}(t) & c_{22}(t) 
\end{array}
\right]\;,
\eeqn
where $c_{ij}(t)=\langle\langle B_{i}^{\pm}(t){\bar 
B}_{j}^{\pm}(0)\rangle\rangle$.
Next we write $C(t)$ in terms of a transfer matrix $\lambda(t,t_{0})$,
%
%
\beqn
{\cal C}(t)\psi=\lambda(t,t_{0}){\cal C}(t_{0})\psi\;,
\eeqn
where
$t_{0}$ is fixed and $t>t_{0}$. If only two states are propagating 
in a given system, the masses of the two states $E_{\alpha}$ ($E_{1}>E_{0}$) 
are given by the eigenvalues of the transfer matrix:
%
%
\beqn
\lambda_{\alpha}(t,t_{0}) = e^{-(t-t_{0})E_{\alpha}}\;\;\;(\alpha 
=0,1)\;,
\eeqn
where $E_{\alpha}$ is independent of $t_{0}$.
The smaller eigenvalue ($\lambda_{1}$) and larger eigenvalue 
($\lambda_{0}$) refer to the masses of the excited state and the 
ground state respectively. 
In general, the system may have more than two states. Thus, we
assume that two states become effectively dominant for an appropriately large 
time-slice $t_{0}$, which can be determined by checking
the sensitivity of $E_{\alpha}$ with respect to variations of $t_{0}$.

We calculate the eigenvalues $\lambda_{\alpha}(t,t_{0})$ of 
${\cal C}(t_{0})^{-1}{\cal C}(t)$ for $t_{0}=3$, and then
evaluate the effective masses of $N$ and $N^\prime$ from
the larger and smaller eigenvalues, respectively, with the statistical errors
coming from the jack-knife method. Using these error estimates,
we determine the
masses in Table VI from weighted averages over the 
time-slice ranges listed there.
Figure 6 shows that the results are quite consistent with the masses
determined from single exponential fits to 
$\langle\langle B_{2}^{+}(t) {\bar B}_{2}^{+}(0)\rangle\rangle$
and
$\langle\langle B_{1}^{+}(t) {\bar B}_{1}^{+}(0)\rangle\rangle$.
We have checked that the variations of $t_{0}$ around $t_{0}=3$ 
do not significantly affect the effective masses.
For $m_{f} \le 0.03$, we unfortunately could not extract the 
mass of the first excited nucleon since a good plateau in the 
effective mass plot is absent.
We note that for $m_f=0.04$ there is still a large splitting in
the eigenvalues, indicating
the overlap of $B_{2}^{+}$ with the nucleon remains small.

In contrast to the positive-parity state, 
the eigenvalues in the negative-parity case appear degenerate,
as shown in Figure 7.
This will be discussed in more detail in the next section.

As we have seen, the first excited state may die out so quickly that
only a few time-slices are available for evaluating the mass,
even if the excited state is well separated from the 
ground state.
To circumvent this quick damping, simulations performed on an anisotropic 
lattice where the temporal lattice spacing is finer than the spatial 
one may be useful. In fact, a recent lattice study~\cite{FXLee2} 
has shown this to be an effective way 
to extract masses of nucleon excited states.

\subsection{Comparison with experiment}
\indent\indent
In the physical spectrum, we have another negative-parity state, $N^*(1650)$,
which is just above the lowest state $N^*(1535)$. 
Although these states are quite
close to each other, they are easily distinguished due to a 
peculiar decay mode of $N^*(1535)$. It is well known that 
the decay rate of $N^*(1535)\rightarrow N+\eta$ is comparable with
$N^*(1535)\rightarrow N+\pi$ even though the $N\pi$ decay mode 
is kinematically favored over $N\eta$. For
the case of $N^*(1650)$ the $N\pi$ decay is dominant;
the branching ratio of $N\eta$
is only a few percent. 
Thus, $N^*(1535)\rightarrow N+\pi$ seems to be
anomalously suppressed. 
However, the corresponding states cannot be easily distinguished 
without knowledge of decay patterns
if they lie close to each other.

As mentioned, for the negative-parity
case our data shows that both eigenvalues of the 
transfer matrix are the same within statistical 
errors (see Figure 7). 
Of course, this does not rule out the possibility that
the splitting might become clear with more statistics and
in the lighter quark mass region. 
That is because the central value of the difference between 
the two eigenvalues increases in the lighter quark mass region.
Furthermore, the value of about 100 MeV at $m_{f}=0.03$ 
seems to be large enough to reproduce experimental splitting in the 
chiral limit.

Nevertheless, unlike the positive-parity case, 
we have no indication of the presence of two 
independent negative-parity states in our calculation where
the mixed correlator yields a consistent mass with both 
$B_{1}^{-}$ and $B_{2}^{-}$ interpolating operators, 
as listed in Table VI.
Thus, our combined results from all data 
for the negative-parity nucleon allows two other possibilities 
regarding $N^*(1650)$. 
One is that $N^*(1535)$ and $N^*(1650)$ are completely degenerate or 
not independent in the quenched calculation.
This situation resembles naive quark models and the MIT bag model
where $N^*(1535)$ and $N^*(1650)$ are degenerate
if we neglect the spin dependent interaction.
In this case, the analysis through the diagonalization of the 2$\times$2 
correlator is no longer helpful.

A Second possibility is that $N^*(1650)$ is simply missing in our
calculation, \ie
both $B_{1}^{-}$ and $B_{2}^{-}$ operators may not couple, or
couple weakly, to the
$N^*(1650)$ state. 
This may be related to an argument in \cite{Jido} where the
authors show that the desired vanishing of the 
phenomenological $\pi NN^*$ coupling
for the $N^*(1535)$ state results simply from the chiral transformation 
properties of particular interpolating operators 
$B_{1}^{-}$ and $B_{2}^{-}$ with the help of the soft pion limit.
If their argument is relevant to the
suppression of the $N\pi$ decay, it is possible that $B_{1}^{-}$ 
and $B_{2}^{-}$ couple strongly to $N^*(1535)$ but not to $N^*(1650)$. 
However, their argument is no longer applicable to
a third interpolating operator 
%
%
\beqn
B_{3\;\mu}^-=(u^{T}C\sigma_{\alpha \beta}u)\sigma_{\alpha 
\beta}\gamma_{\mu}d-(u^{T}C\sigma_{\alpha \beta}d)\sigma_{\alpha 
\beta}\gamma_{\mu}u\;,
\eeqn
which belongs to the chiral multiplet $(\frac{1}{2},1)\oplus(1,\frac{1}{2})$
and has no derivative~\cite{Ioffe,Cohen}. The $B_{3\;\mu}^-$
couples to both $J=3/2$ and $J=1/2$ states.
In this case, $N^*(1650)$ might be extracted 
from the $B_{3\;\mu}^-$ operator since there is no reason for 
lack of coupling to the $N^*(1650)$ state.  

Finally, we would like to mention a remaining puzzle. We have 
done the first successful lattice calculation of both $N^*$ and $N'$ spectra.
As for the $N'$, we have reliable data only for relatively large 
values of $m_{f}$. Comparing the $N'$ mass with the $N^*$ mass, we 
find that the ordering of $N'$ and $N^*$ is inverted compared to experiment.
Furthermore, the level spacing
between $N$-$N^*$ and $N^*$-$N'$ is almost even~\cite{Sasaki}. 
What we see here
closely resembles the wrong ordering problem of the excited 
nucleon spectrum in naive quark models and the MIT bag model.
However, if this result is true for heavy quarks, we can make
an important prediction:
the first excited state of the spin one-half $\Sigma_{c}$ is
a negative-parity state rather than a positive-parity state.

Unfortunately, our statistics did not allow the computation of the
$N^\prime$ mass in the light-quark region ($m_{f}<0.03$).
In addition, our data show no evidence for
the possibility of switching the level ordering between $N^*$ and $N'$ 
towards the chiral limit. Indeed, we did not observe that 
the $N'$ mass decreased faster than the $N^*$ mass 
with decreasing quark mass. However, finite volume effects for 
higher excited states may be more serious than for lower excited states.
According to naive quark models or the MIT bag model,
the $N'$ state is radially excited in contrast to the nucleon ground state 
and the $N^*$ state. We note that the trend in the three heaviest
quark mass points is at least consistent with the switching of the
ordering in the chiral limit.
Needless to say, the quenched approximation
may also play a role in this puzzle.
In summary, our results do not rule out the possibility of switching
the ordering between $N^*$ and $N'$ near the chiral limit,
and to solve this remaining puzzle we need further 
systematic calculations toward the chiral limit.

\section{Conclusion}
\indent\indent
We have studied the mass spectrum of the parity partner of the nucleon in
quenched lattice QCD using domain wall fermions which preserve
chiral symmetry to a high degree in lattice simulations. 
Most importantly we demonstrated that this method is capable of
calculating the mass of the negative-parity $N^*$ state in the 
spin one-half and isospin one-half baryon sector.

We made a systematic investigation of the $N^*$ spectrum by using two
distinct interpolating operators, $B_{1}^-$ and $B_{2}^-$.  We found the
$N^*$ mass estimates extracted from them agree with each other.  In
practice the $B_{1}^-$ correlator is more convenient than $B_{2}^-$
in extracting the $N^*$ mass because it is 
less noisy, especially in the light quark mass region.

We have found a definite mass splitting 
between $N$ and $N^*$ states in the whole quark
mass range we studied. Furthermore, this splitting grows with decreasing
quark mass. The $N^*$ mass and the $N$-$N^*$
mass splitting in the chiral limit obtained by extrapolation are
consistent with the experimental value within about 5-10\%, depending
on the mass we choose to set the lattice spacing, the nucleon or 
the $\rho$ meson. These results have been confirmed by subsequent
lattice calculations using improved Wilson 
fermions~\cite{FXLee2,Richards2}.
Needless to say this is very encouraging for further investigations of
$N^*$ physics using lattice QCD simulations.

In contrast to the negative parity operators, the positive parity
operators, $B_1^+$ and $B_2^+$, yield distinct mass
signals.  From the $B_1^+$ operator we obtained a clean signal for the 
nucleon ground state, while from $B_2^+$ we always found a 
heavier mass.  This is probably because the latter vanishes 
in the non-relativistic limit and 
thus has small overlap with the ground state.
Indeed, we confirmed numerically that
$\langle 0 | B_ 2^+ | N \rangle \simeq 0$ and that
$B_{2}^{+}$ yields the mass signal of the positive-parity
excited state in the
heavy quark region. We found
that in the heavy quark mass region the mass of this excited state is
about twice as high above the ground state mass as the 
$N^*$ mass.
This property of the 
$B_2^+$ operator was confirmed by comparing the results with the
excited-state spectrum obtained from the diagonalization of a 2$\times$2
correlation matrix constructed from both interpolating operators and
also by examining single particle fits to the mixed correlation function.

We did not resolve a long-standing puzzle regarding 
the excited-state spectrum of the nucleon, the inversion of the
positive- and negative- parity first excited states with respect to
experiment. However, we believe that the level switching 
between $N^*$ and $N'$ 
must occur close to the chiral limit, though there is scant evidence
in our results that this might happen.
Taking our calculation at face value
leads to the prediction that the first excited state of 
$\Sigma_c$ has $J^P=1/2^-$.

Needless to say, more work needs to be accomplished in order to achieve a
fully systematic calculation of the excited nucleon spectrum. In the
near future we plan to increase
our volume and statistics to further explore the chiral limit, and to check 
that lattice spacing errors are small by running at several gauge couplings. We
expect that the latter effect is small due to the chiral symmetry properties of domain
wall fermions. A long term goal is to include dynamical fermions.

\section*{Acknowledgment}

The authors acknowledge useful discussions with all 
the members of the RIKEN-BNL-Columbia-KEK collaboration, but especially N.
Christ, R. 
Mawhinney and A. Soni. All lattice simulations in this paper were done 
on the RIKEN BNL QCDSP supercomputer
as part of the RIKEN-BNL-Columbia-KEK collaboration.
We thank RIKEN, Brookhaven National
Laboratory, and the U.S. Department of Energy for providing the facilities 
essential for the completion of this work. Finally, S.S. 
thanks all of the members of the RIKEN-BNL Research Center for 
their warm hospitality during his stay, where most of the 
present study has been carried out.

\newpage

\section*{Appendix A: General chiral transformation and parity 
partners}

\indent\indent
To extend our discussion of Sec.III-C to general chiral 
transformations, we prepare iso-doublet operators
for $B^{+}_{1}$ and $B^{+}_{2}$,
%
%
\barr
{\cal B}_{1}^{+}(x)
&=&
\left(
\begin{array}{r}
\varepsilon_{abc}[u^{T}_{a}(x)C\gamma_{5}d_{b}(x)]u_{c}(x) \\
\varepsilon_{abc}[d^{T}_{a}(x)C\gamma_{5}u_{b}(x)]d_{c}(x)  
\end{array}
\right)\;, \\
{\cal B}_{2}^{+}(x)
&=&
\left(
\begin{array}{r}
\varepsilon_{abc}[u^{T}_{a}(x)Cd_{b}(x)]\gamma_{5}u_{c}(x) \\
\varepsilon_{abc}[d^{T}_{a}(x)Cu_{b}(x)]\gamma_{5}d_{c}(x)  
\end{array}
\right)\;.
\earr
The upper component corresponds to the proton and the lower component 
corresponds to the neutron. Also we can define iso-doublet 
operators for the negative parity nucleons as ${\cal 
B}^{-}_{1,2}=\gamma_{5}{\cal B}^{+}_{1,2}$, respectively.
Here we consider a general transformation
of $SU(2)_{V}$ and $SU(2)_{A}$ symmetry,
%
%
\barr
\left[\right.{\cal Q}_{\small V}^{a}, 
q(x)\left]\right.&=&i\tau_{a}q(x)\;, \\
\left[\right.{\cal Q}_{\small A}^{a}, q(x)\left]\right.&=&i\tau_{a}\gamma_{5}
q(x)\;, 
\earr
where $q=(u,d)^{T}$ and $\tau_{a}$ is a Pauli matrix.
Under these transformations, the two-point correlator composed of 
either ${\cal B}_{1}^{\pm}$ or ${\cal B}_{2}^{\pm}$
should transform as
%
%
\barr
\left[ {\cal Q}_{\small V}^{a}, {\cal B}^{\pm}_{1,2}(x)
{\bar {\cal B}}^{\pm}_{1,2}(0)\right]
&=&i\left[\tau_{a}, {\cal B}^{\pm}_{1,2}(x)
{\bar {\cal B}}^{\pm}_{1,2}(0)\right] \;,\\
\label{eq:vector}
\left[ {\cal Q}_{\small A}^{a}, {\cal B}^{\pm}_{1,2}(x)
{\bar {\cal B}}^{\pm}_{1,2}(0)\right]
&=&
i\left\{\gamma_{5}\tau_{a}, {\cal B}^{\pm}_{1,2}(x)
{\bar {\cal B}}^{\pm}_{1,2}(0)\right\} \;.
\label{eq:axial}
\earr
Eq.(\ref{eq:vector}) along 
with the fact that ${\cal Q}_{\small V}^{a}\rvac = 0$ tells us
that the nucleon two-point correlation function 
should be commute with $\tau_{a}$,
%
%
\beqn
\left[\tau_{a}, G_{\cal B^{\pm}}(t)\right]=0\;,
\eeqn
which means that $G_{\cal B^{+}}(t)$ has only diagonal elements of
iso-spin indices. In other words, the proton state and the 
neutron state are eigenstates of iso-spin, of course.
If ${\cal Q}_{\small A}^{a}\rvac = 0$, it turns out that

%
%
\beqn
\left\{\gamma_{5}\tau_{a}, G_{\cal B^{\pm}}(t)\right\}
=
\tau_{a}\left\{\gamma_{5}, G_{\cal B^{\pm}}(t) \right\}=0\;.
\eeqn
Thus, we can obtain the strict relation in terms of two-point correlators 
between the nucleon and its parity partner
%
%
\beqn
G_{\cal B^{+}}(t) = G_{\cal B^{-}}(t)
\eeqn
in the non-broken phase of chiral symmetry.

\newpage
\centerline{\large FIGURE CAPTIONS}
\begin{description}

\vspace{1.0cm}
\item[FIG.1A]
\begin{minipage}[t]{13cm}
\baselineskip=20pt
The effective mass of the nucleon
from the $\langle\langle B_{1}^{+}(t){\bar B}_{1}^{+}(0)\rangle\rangle$ 
correlator ($\times$) and its parity partner 
from $\langle\langle B_{1}^{-}(t){\bar B}_{1}^{-}(0)\rangle\rangle$ 
($\Box$) 
and $\langle\langle B_{2}^{-}(t){\bar B}_{2}^{-}(0)\rangle\rangle$ 
($\Diamond$) correlators on an ensemble of 405 configurations at 
$m_{f}=0.03$. 
The solid lines and dashed lines represent 
each fitted mass and its statistical error.
\end{minipage}

\vspace{1.0cm}
\item[FIG.1B]
\begin{minipage}[t]{13cm}
\baselineskip=20pt
The effective mass of the nucleon
from the $\langle\langle B_{1}^{+}(t){\bar B}_{1}^{+}(0)\rangle\rangle$ 
correlator ($\times$) and its parity partner 
from $\langle\langle B_{1}^{-}(t){\bar B}_{1}^{-}(0)\rangle\rangle$ 
($\Box$) 
and $\langle\langle B_{2}^{-}(t){\bar B}_{2}^{-}(0)\rangle\rangle$ 
($\Diamond$) correlators on an ensemble of 305 configurations at 
$m_{f}=0.05$. 
The solid lines and dashed lines represent 
each fitted mass and its statistical error.
\end{minipage}

\vspace{1.0cm}
\item[FIG.1C]
\begin{minipage}[t]{13cm}
\baselineskip=20pt
The effective mass of the nucleon
from the $\langle\langle B_{1}^{+}(t){\bar B}_{1}^{+}(0)\rangle\rangle$ 
correlator ($\times$) and its parity partner 
from $\langle\langle B_{1}^{-}(t){\bar B}_{1}^{-}(0)\rangle\rangle$ 
($\Box$) 
and $\langle\langle B_{2}^{-}(t){\bar B}_{2}^{-}(0)\rangle\rangle$ 
($\Diamond$) correlators on an ensemble of 105 configurations at 
$m_{f}=0.10$. 
The solid lines and dashed lines represent 
each fitted mass and its statistical error.
\end{minipage}

\vspace{1.0cm}
\item[FIG.2.]
\begin{minipage}[t]{13cm}
\baselineskip=20pt
$N$ ($\times$) and $N^*$ ($\Box$ and $\Diamond$) masses versus the quark mass
$m_{f}$ in lattice units ($a^{-1}\approx 1.9$GeV from $aM_{\rho}$=0.400(8) in
the chiral limit). The corresponding experimental values for $N$ and 
$N^*$ are marked with lower and upper stars. 
The $N$-$N^*$ mass splitting is clearly observed.
Symbols ($\times$, $\Box$ and $\Diamond$) are defined as in Figure 1.

\end{minipage}

\vspace{1.0cm}
\item[FIG.3.]
\begin{minipage}[t]{13cm}
\baselineskip=20pt
Mass ratio of the negative-parity excited-state and positive-parity
ground-state 
baryons versus mass ratio of the pseudoscalar meson and vector meson.
All calculations are done for three degenerate valence-quarks.
Ratios are calculated using
fitted masses from the 
$\langle\langle B_{1}^{-}(t){\bar B}_{1}^{-}(0)\rangle\rangle$ 
($\Box$) 
and $\langle\langle B_{2}^{-}(t){\bar B}_{2}^{-}(0)\rangle\rangle$ 
($\Diamond$) correlators. Experimental points are denoted by stars.
\end{minipage}

\vspace{1.0cm}
\item[FIG.4A]
\begin{minipage}[t]{13cm}
\baselineskip=20pt
The effective mass from the
$\langle\langle B_{1}^{+}(t){\bar B}_{1}^{+}(0)\rangle\rangle$ 
($\times$) and
$\langle\langle B_{2}^{+}(t){\bar B}_{2}^{+}(0)\rangle\rangle$ 
($\circ$) correlators on an ensemble of 405 configurations at $m_{f}=0.03$.
The solid lines and dashed lines represent 
each fitted mass and its statistical error.
\end{minipage}

\vspace{1.0cm}
\item[FIG.4B]
\begin{minipage}[t]{13cm}
\baselineskip=20pt
The effective mass from the
$\langle\langle B_{1}^{+}(t){\bar B}_{1}^{+}(0)\rangle\rangle$ 
($\times$) and
$\langle\langle B_{2}^{+}(t){\bar B}_{2}^{+}(0)\rangle\rangle$ 
($\circ$) correlators on an ensemble of 305 configurations at $m_{f}=0.05$.
The solid lines and dashed lines represent 
each fitted mass and its statistical error.
\end{minipage}

\vspace{1.0cm}
\item[FIG.4C]
\begin{minipage}[t]{13cm}
\baselineskip=20pt
The effective mass from the
$\langle\langle B_{1}^{+}(t){\bar B}_{1}^{+}(0)\rangle\rangle$ 
($\times$) and
$\langle\langle B_{2}^{+}(t){\bar B}_{2}^{+}(0)\rangle\rangle$ 
($\circ$) correlators on an ensemble of 105 configurations at 
$m_{f}=0.10$.
The solid lines and dashed lines represent 
each fitted mass and its statistical error.
\end{minipage}

\vspace{1.0cm}
\item[FIG.5.]
\begin{minipage}[t]{13cm}
\baselineskip=20pt
The fitted mass from
$\langle\langle B_{1}^{+}(t){\bar B}_{1}^{+}(0)\rangle\rangle$ 
($\times$),
$\langle\langle B_{2}^{+}(t){\bar B}_{2}^{+}(0)\rangle\rangle$ 
($\circ$) and mixed type
$\langle\langle B_{1}^{+}(t){\bar B}_{2}^{+}(0) 
+ B_{2}^{+}(t){\bar B}_{1}^{+}(0)\rangle\rangle$ 
($\Diamond$) correlators.
The corresponding experimental values for $N$ and 
$N'$ are marked with lower and upper stars. Note, the values extracted
from the mixed correlation
function do not represent the mass of an actual 
asymptotic state, except possibly
in the light and heavy quark limits.
\end{minipage}

\vspace{1.0cm}
\item[FIG.6.]
\begin{minipage}[t]{13cm}
\baselineskip=20pt
Comparison of
the fitted mass from $\langle\langle B_{2}^{+}(t){\bar B}_{2}^{+}(0)
\rangle\rangle$ ($\circ$) 
and the estimated mass from the average effective mass of
the smaller eigenvalue of the transfer matrix ($\bullet$).
The symbol $\times$ corresponds to the nucleon ground-state mass evaluated from
the larger eigenvalue of the transfer matrix, which is quite
consistent with the fitted mass from $\langle\langle B_{1}^{+}(t)
{\bar B}_{1}^{+}(0)\rangle\rangle$.
\end{minipage}

\vspace{1.0cm}
\item[FIG.7.]
\begin{minipage}[t]{13cm}
\baselineskip=20pt
Symbols $\Box$ and $\Diamond$ correspond to the the estimated mass from 
the average effective mass of the larger and smaller eigenvalues
of the transfer matrix for the negative parity state. The eigenvalues 
are degenerate within errors. Symbols $\times$ and $\circ$ 
are defined as in Figure 6. The corresponding experimental values
for $N(939)$, $N'(1440)$ and $N^*(1535)$ are marked with
lower, middle and upper stars.
The ordering of the negative-parity nucleon
($N^*$) and the positive-parity excited nucleon ($N'$) 
is inverted relative to experiment.
\end{minipage}
\end{description}

\newpage
%
%
\begin{table}
\begin{center}
\begin{tabular}{ccccc}
\hline
\hline
$am_{f}$ & $aM_{\pi}$ & $aM_{\rho}$ & $M_{\pi}/M_{\rho}$ & \# of configs.
\\ \hline
 0.020 &  0.2687(24) &  0.4530(62) & 0.593(14) & 98\\
 0.030 &  0.3224(21) &  0.4814(45) & 0.670(11) & 98\\
 0.040 &  0.3691(19) &  0.5126(42) & 0.720(10) & 98\\
 0.050 &  0.4116(18) &  0.5395(36) & 0.763(8)  & 98\\
 0.075 &  0.5080(17) &  0.6088(38) & 0.834(8)  & 98\\
 0.100 &  0.5962(18) &  0.6797(36) & 0.877(7)  & 98\\
 0.125 &  0.6774(17) &  0.7483(33) & 0.905(6)  & 98\\ 
\hline
\hline
\end{tabular}
\end{center}
\caption{Single exponential fit of $\pi$ and $\rho$ two-point correlators at 
$\beta=6.0$ on a $16^{3}\times 32 \times 16$ lattice with $M_{5}=1.8$.
The masses for $am_f<0.05$ agree within errors with those given
in Ref.~\protect\cite{RBC1} 
which were calculated from different correlation
functions on a slightly different set of configurations.}
\label{tab:massratios}
\end{table}

%
%
%
\begin{table}
\begin{center}
\begin{tabular}{c|clccccc}
\hline
\hline
state ($J^P$) & $am_{f}$ & $aM_{N}$(stat.)(sys.) &  $\chi^2/N_{\rm DF}$ & $t_{\rm min}$
& $t_{\rm max}$ &
conf. level & \# of configs. \\ \hline
$N$ ($1/2^+$)
&0.020 & 0.654(12)(14)& 1.00  & 12  & 20  & 0.43 & 405 \\
&0.030 & 0.716(5)(0) & 1.09  & 10  & 20  & 0.37 & 405 \\
&0.040 & 0.754(6)(1) & 1.08  & 11  & 20  & 0.38 & 305 \\
&0.050 & 0.805(5)(2) & 1.17  & 11  & 20  & 0.31 & 305 \\
&0.075 & 0.929(7)(4) & 0.81  & 12  & 20  & 0.58 & 105 \\
&0.100 & 1.045(5)(3) & 0.94  & 12  & 20  & 0.47 & 105 \\
&0.125 & 1.162(6)(4) & 0.84  & 15  & 20  & 0.50 & 105 \\
\hline
\hline
\end{tabular}
\end{center}
\caption{Single exponential fit of the nucleon two-point correlator
$\langle\langle B_{1}^{+}(t){\bar B}_{1}^{+}(0)\rangle\rangle$ at 
$\beta=6.0$ on a $16^{3}\times 32 \times 16$ lattice with $M_{5}=1.8$.
The systematic error is estimated from the change in the fitted
mass when $t_{\rm min}$ is increased by one.
The masses for $am_f<0.05$ agree within errors with those given 
in Ref.~\protect\cite{RBC1} which were calculated from different correlation 
functions on a slightly different set of configurations.}
\label{tab:fit1}
\end{table}
%

%
%
\begin{table}
\begin{center}
\begin{tabular}{c|clccccc}
\hline
\hline
state ($J^P$) & $am_{f}$ & $aM_{N^*}$(stat.)(sys.) & $\chi^2/N_{\rm DF}$ & $t_{\rm min}$
& $t_{\rm max}$ &
conf. level & \# of configs. \\ \hline
$N^*$ ($1/2^-$)
&0.020 & 0.935(34)(40)  & 1.29  &  6  & 10  & 0.28 & 405 \\
&0.030 & 0.959(24)(24)  & 1.16  &  7  & 12  & 0.33 & 405 \\
&0.040 & 1.018(28)(56)  & 1.49  &  8  & 12  & 0.22 & 305 \\
&0.050 & 1.048(20)(38)  & 1.26  &  8  & 13  & 0.28 & 305 \\
&0.075 & 1.104(13)(18)  & 1.27  &  7  & 13  & 0.27 & 105 \\
&0.100 & 1.203(10)(12)  & 0.99  &  7  & 12  & 0.41 & 105 \\
&0.125 & 1.303(8)(8)    & 0.92  &  7  & 12  & 0.45 & 105 \\
\hline
\hline
\end{tabular}
\end{center}
\caption{Single exponential fit of the negative-parity nucleon two-point 
correlator $\langle\langle B_{1}^{-}(t){\bar B}_{1}^{-}(0)\rangle\rangle$ at 
$\beta=6.0$ on a $16^{3}\times 32 \times 16$ lattice with $M_{5}=1.8$.
The systematic error is estimated from the change in the fitted
mass when $t_{\rm min}$ is increased by one.}
\label{tab:fit2}
\end{table}
%

%
%
\begin{table}
\begin{center}
\begin{tabular}{c|clccccc}
\hline
\hline
state ($J^P$) & $am_{f}$ & $aM_{N^*}$(stat.)(sys.) & $\chi^2/N_{\rm DF}$ & $t_{\rm min}$
& $t_{\rm max}$ &
conf. level & \# of configs. \\ \hline
$N^*$ ($1/2^-$)
&0.030 & 0.982(21)(15)  & 1.10  &  5  & 10  & 0.35 & 405 \\
&0.040 & 0.993(16)(17)  & 0.81  &  5  & 10  & 0.52 & 305 \\
&0.050 & 1.029(15)(16)  & 1.00  &  6  & 10  & 0.39 & 305 \\
&0.075 & 1.092(12)(10)  & 0.75  &  5  & 11  & 0.59 & 105 \\
&0.100 & 1.193(11)(16)  & 0.82  &  6  & 12  & 0.54 & 105 \\
&0.125 & 1.308(13)(9)   & 0.84  &  7  & 15  & 0.55 & 105 \\
\hline
\hline
\end{tabular}
\end{center}
\caption{Single exponential fit of the negative-parity nucleon two-point
correlator
$\langle\langle B_{2}^{-}(t){\bar B}_{2}^{-}(0)\rangle\rangle$ at 
$\beta=6.0$ on a $16^{3}\times 32 \times 16$ lattice with $M_{5}=1.8$.
The systematic error is estimated from the change in the fitted
mass when $t_{\rm min}$ is increased by one.}
\label{tab:fit3}
\end{table}
%

%
%
\begin{table}
\begin{center}
\begin{tabular}{c|clccccc}
\hline
\hline
state ($J^P$) & $am_{f}$ & $aM_{N'}$(stat.)(sys.) & $\chi^2/N_{\rm DF}$ & $t_{\rm min}$
& $t_{\rm max}$ &
conf. level & \# of configs. \\ \hline
$N'$ ($1/2^+$)
&0.030 & 1.170(50)(108)  & 1.47  &  5  &  8  & 0.23 & 405 \\
&0.040 & 1.191(36)(69)   & 1.28  &  5  & 11  & 0.27 & 305 \\
&0.050 & 1.244(43)(148)  & 1.49  &  6  & 11  & 0.20 & 305 \\
&0.075 & 1.303(39)(58)   & 0.43  &  6  & 10  & 0.73 & 105 \\
&0.100 & 1.387(19)(2)    & 0.63  &  5  & 10  & 0.64 & 105 \\
&0.125 & 1.484(16)(6)    & 1.00  &  5  & 11  & 0.42 & 105 \\
\hline
\hline
\end{tabular}
\end{center}
\caption{Single exponential fit of the unconventional nucleon two-point
correlator
$\langle\langle B_{2}^{+}(t){\bar B}_{2}^{+}(0)\rangle\rangle$ at 
$\beta=6.0$ on a $16^{3}\times 32 \times 16$ lattice with $M_{5}=1.8$.
The systematic error is estimated from the change in the fitted
mass when $t_{\rm min}$ is increased by one.}
\end{table}
%

%
%
\begin{table}
\begin{center}
\begin{tabular}{cc|cc|cc|cc|c}
\hline
\hline
$J^P$ & $am_{f}$ & Mixed type & [$t_{\rm min}, t_{\rm max}$]
& $E_{0}$ & [$t_{\rm min}, t_{\rm max}$] 
& $E_{1}$ & [$t_{\rm min}, t_{\rm max}$] & \# of configs. \\ \hline
$1/2^+$
&0.020 & 0.704(67)  & [6,11] & 0.649(10)  & [ 6,13] & N/A        &       & 200\\
&0.030 & 0.795(57)  & [5,12] & 0.703(9)   & [ 7,15] & N/A        &       & 200\\
&0.040 & 0.873(56)  & [5,12] & 0.755(9)   & [ 7,20] & 1.264(84)  & [5, 8]& 200\\
&0.050 & 0.942(54)  & [5,11] & 0.808(8)   & [ 8,20] & 1.247(67)  & [5,10]& 200\\
&0.075 & 1.198(115) & [5,10] & 0.927(11)  & [11,20] & 1.256(39)  & [4,10]&  81\\
&0.100 & 1.405(54)  & [3,10] & 1.048(9)   & [12,20] & 1.382(39)  & [5,10]&  81\\
&0.125 & 1.504(83)  & [4,10] & 1.161(7)   & [13,20] & 1.479(33)  & [5,10]&  81\\
\hline
$1/2^-$
&0.020 & N/A         &        &0.901(50) & [5,10]  & N/A       &       & 200\\
&0.030 & 0.992(47)   & [5,11] &0.938(29) & [5,10]  & 1.004(49) & [4,10]& 200\\
&0.040 & 0.979(24)   & [5,11] &0.996(27) & [6,10]  & 1.021(35) & [5,10]& 200\\
&0.050 & 1.005(16)   & [5,11] &1.054(29) & [7,11]  & 1.042(26) & [5,11]& 200\\
&0.075 & 1.092(14)   & [5,12] &1.105(18) & [5,10]  & 1.102(22) & [5,10]&  81\\
&0.100 & 1.185(10)   & [5,10] &1.197(18) & [5,12]  & 1.203(16) & [5,12]&  81\\
&0.125 & 1.282(9)    & [5,11] &1.292(16) & [5,12]  & 1.303(13) & [5,12]&  81\\
\hline
\hline
\end{tabular}
\end{center}
\caption{The third column lists the fitted masses of the mixed type correlators. 
Fifth and seventh columns list the weighted averages of the effective mass 
derived from the larger eigenvalue and the smaller eigenvalue of the
transfer matrix 
$\lambda_{a}(t,t_{0}=3)$ induced by the 2$\times$2 matrix correlator. 
The fits corresponding to column three all have $\chi^2/N_{DF}<1.5$
For the positive-parity state, the 
larger and smaller eigenvalue of the transfer matrix are clearly 
distinguishable. They correspond to the nucleon ground state ($N$) and
the first excited nucleon ($N'$) respectively. However, for
the negative parity state, both eigenvalues are degenerate within errors. 
}
\end{table}

\newpage
\centerline{\Large FIG.1a (Phys.Rev.D) Shoichi Sasaki \etal}

\vspace{2.5cm}
%
%
\noindent
\centerline{\epsfxsize=15.0cm
\epsfbox{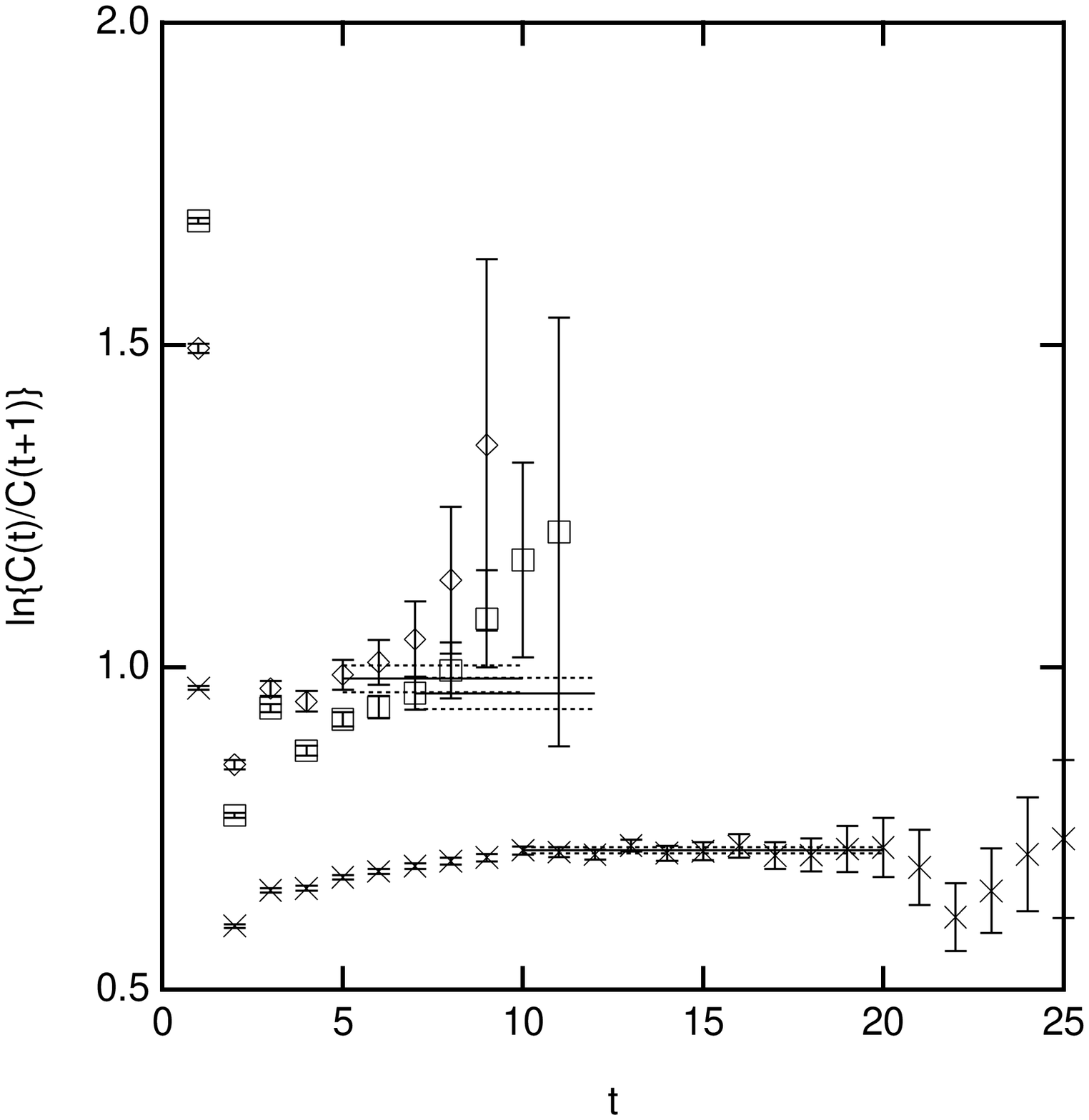}}
\vspace{0.5cm}

\newpage
\centerline{\Large FIG.1b (Phys.Rev.D) Shoichi Sasaki \etal}

\vspace{2.5cm}
%
%
\noindent
\centerline{\epsfxsize=15.0cm
\epsfbox{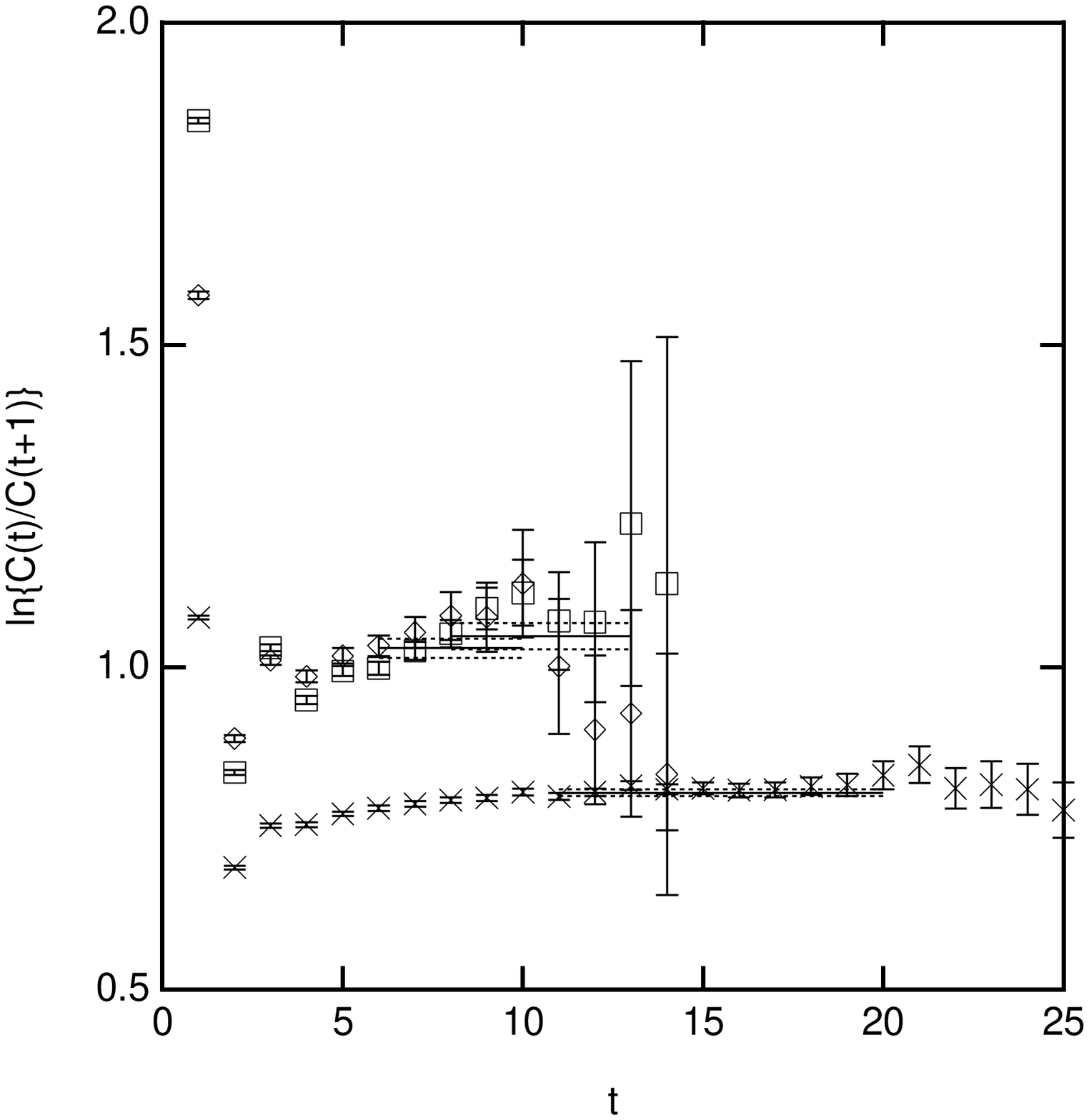}}
\vspace{0.5cm}

\newpage
\centerline{\Large FIG.1c (Phys.Rev.D) Shoichi Sasaki \etal}

\vspace{2.5cm}
%
%
\noindent
\centerline{\epsfxsize=15.0cm
\epsfbox{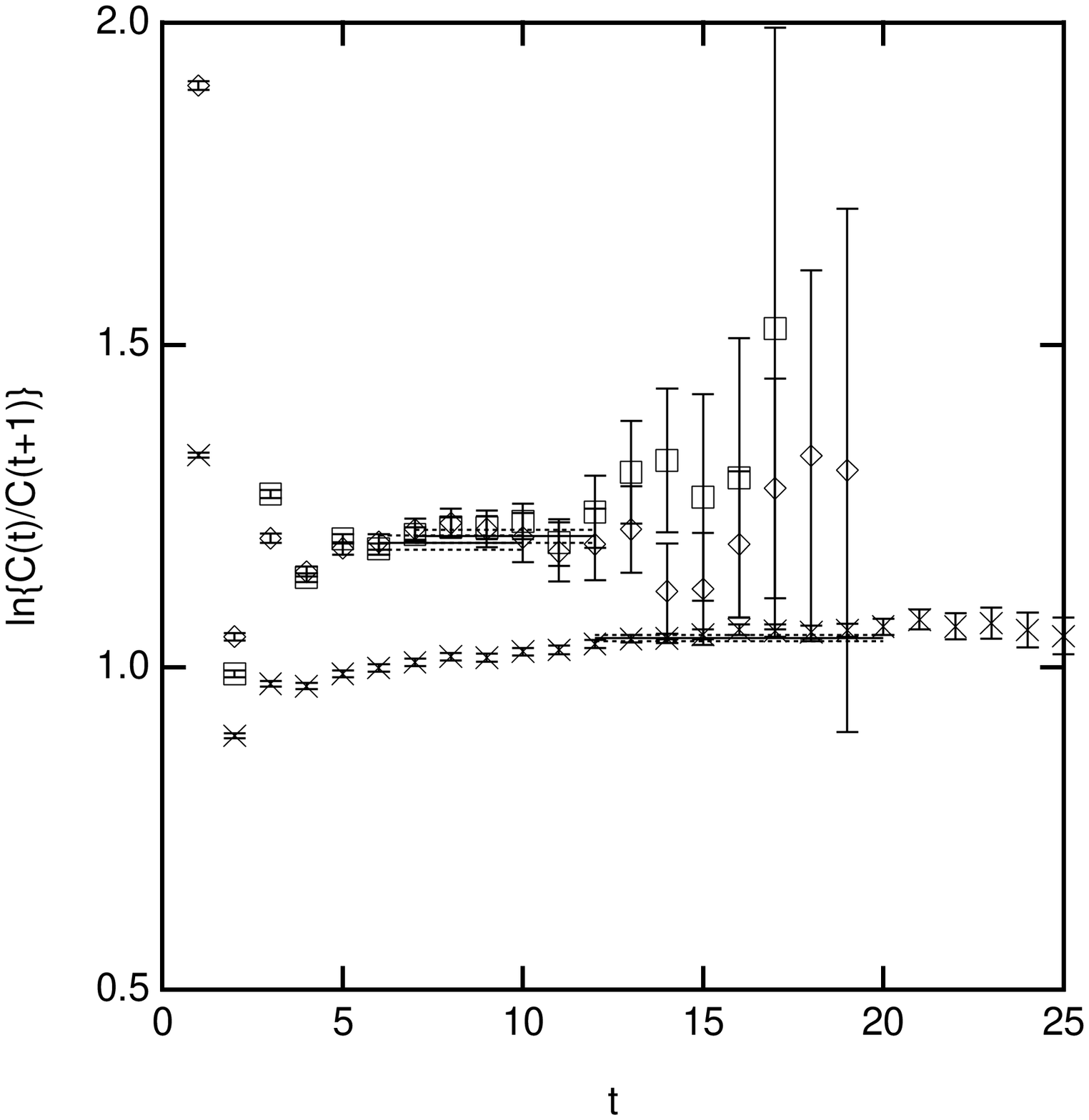}}
\vspace{0.5cm}

\newpage
\centerline{\Large FIG.2 (Phys.Rev.D) Shoichi Sasaki \etal}

\vspace{2.5cm}
%
%
\noindent
\centerline{\epsfxsize=15.0cm
\epsfbox{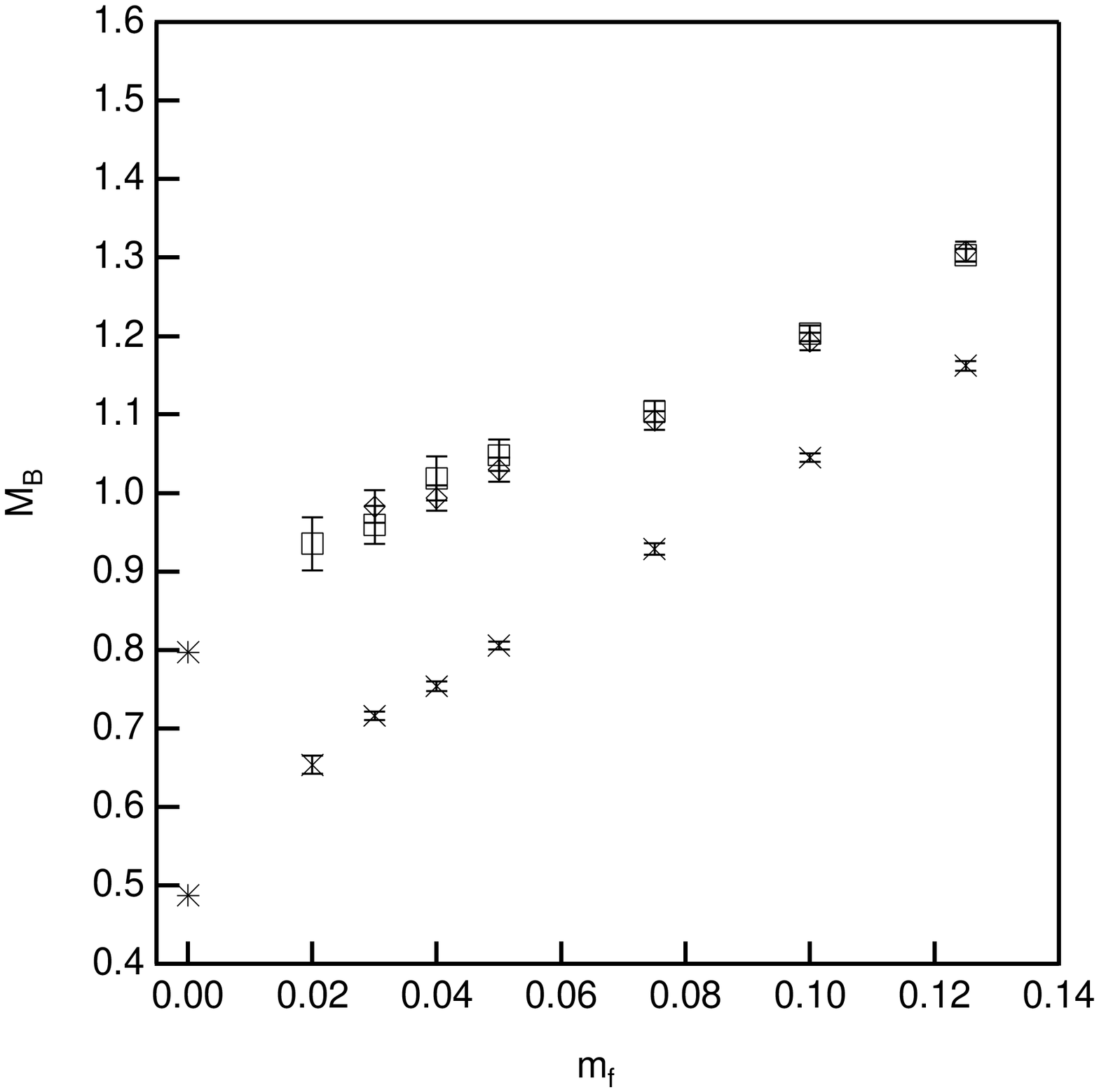}}
\vspace{0.5cm}

\newpage
\centerline{\Large FIG.3 (Phys.Rev.D) Shoichi Sasaki \etal}

\vspace{2.5cm}
%
%
\noindent
\centerline{\epsfxsize=15.0cm
\epsfbox{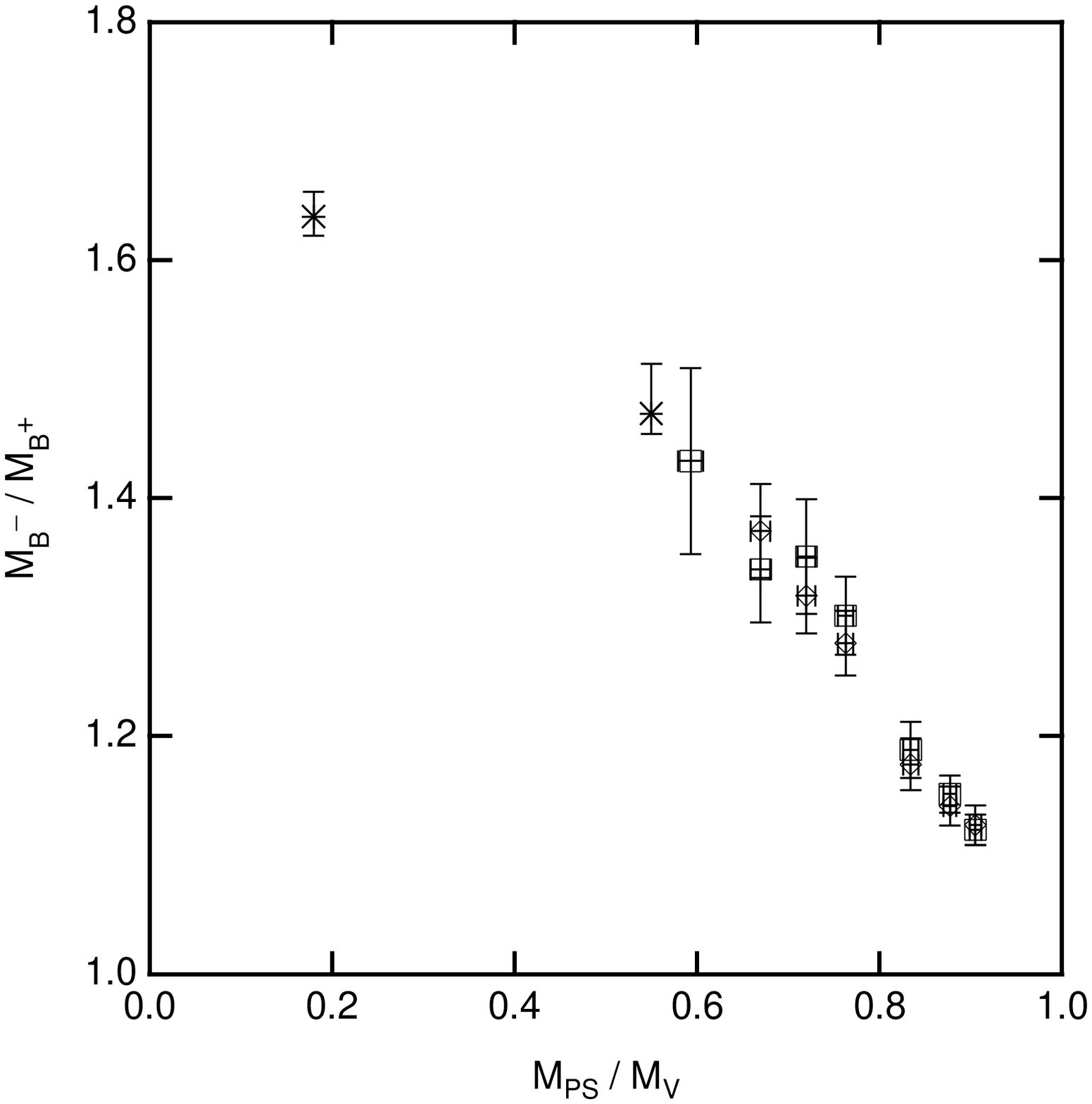}}
\vspace{0.5cm}

\newpage
\centerline{\Large FIG.4a (Phys.Rev.D) Shoichi Sasaki \etal}

\vspace{2.5cm}
%
%
\noindent
\centerline{\epsfxsize=15.0cm
\epsfbox{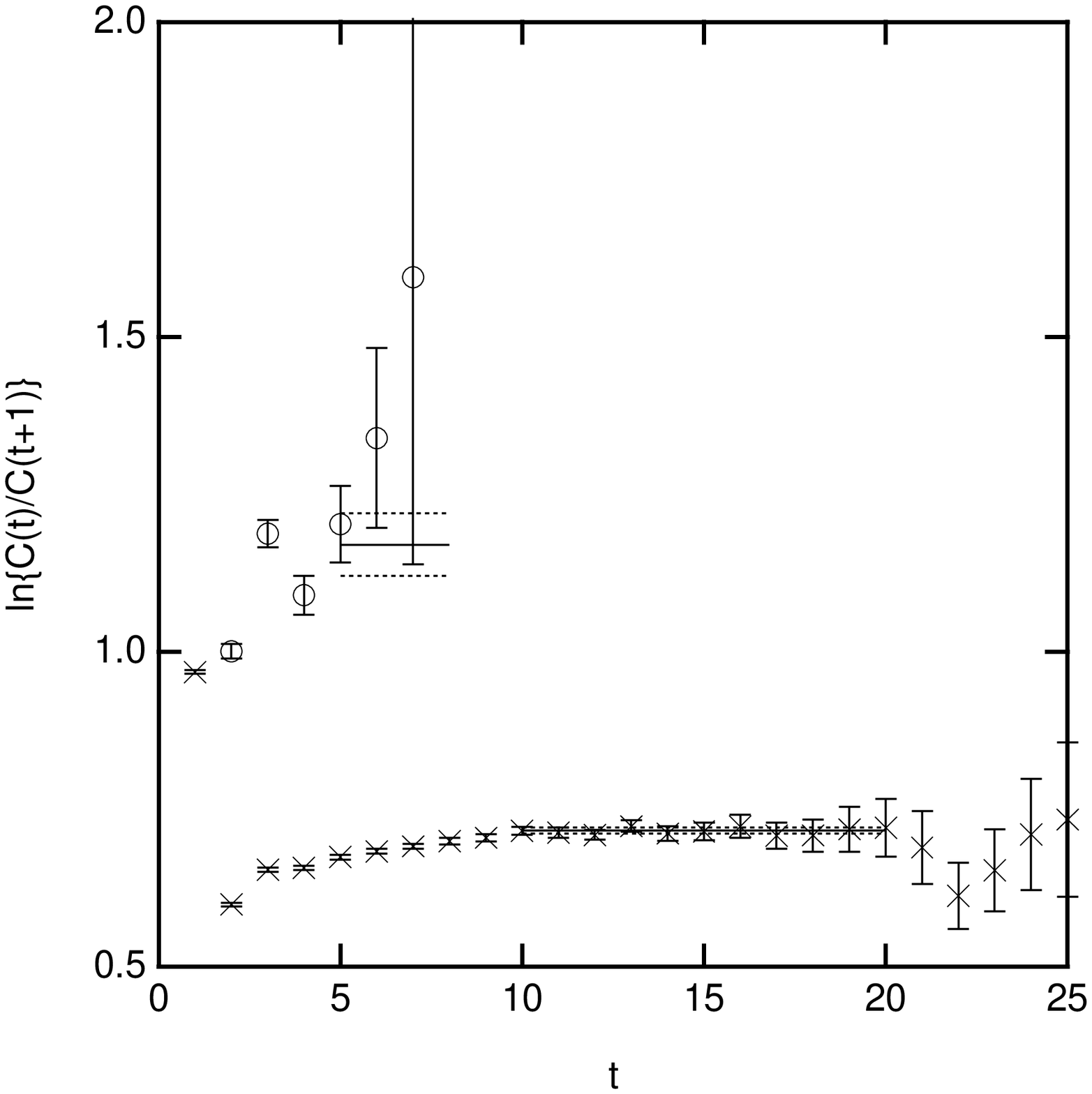}}
\vspace{0.5cm}

\newpage
\centerline{\Large FIG.4b (Phys.Rev.D) Shoichi Sasaki \etal}

\vspace{2.5cm}
%
%
\noindent
\centerline{\epsfxsize=15.0cm
\epsfbox{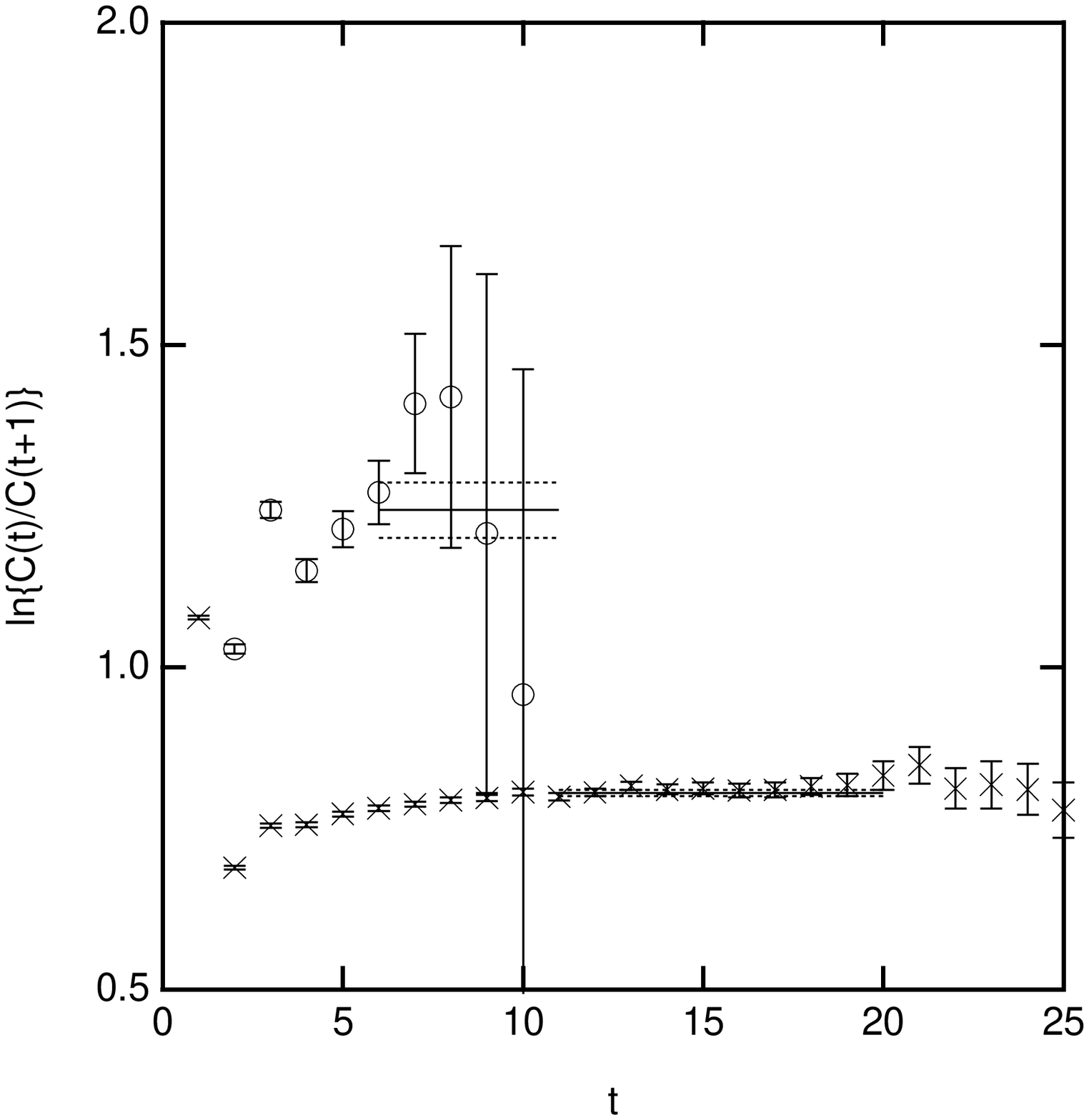}}
\vspace{0.5cm}

\newpage
\centerline{\Large FIG.4c (Phys.Rev.D) Shoichi Sasaki \etal}

\vspace{2.5cm}
%
%
\noindent
\centerline{\epsfxsize=15.0cm
\epsfbox{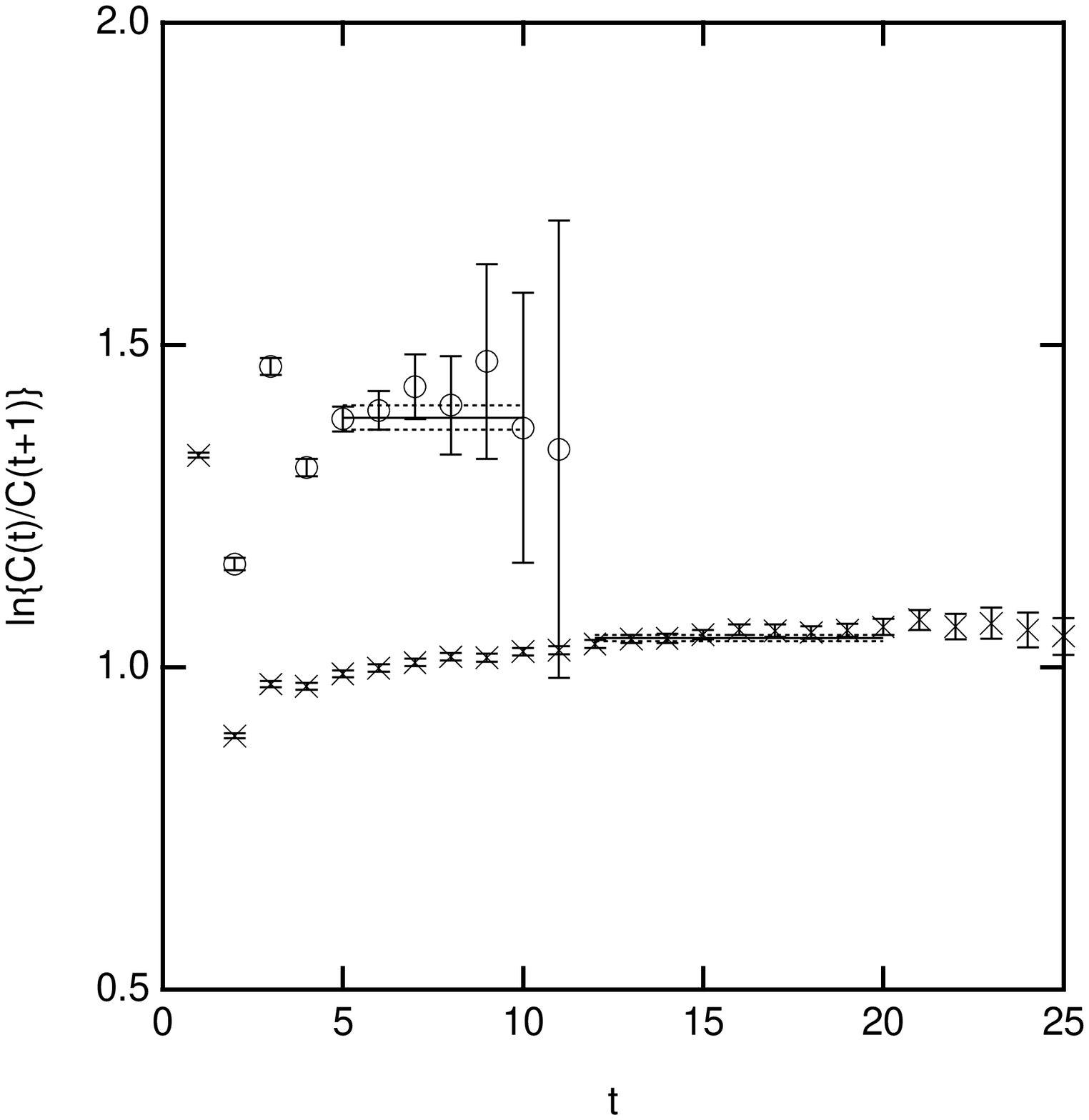}}
\vspace{0.5cm}

\newpage
\centerline{\Large FIG.5 (Phys.Rev.D) Shoichi Sasaki \etal}

\vspace{2.5cm}
%
%
\noindent
\centerline{\epsfxsize=15.0cm
\epsfbox{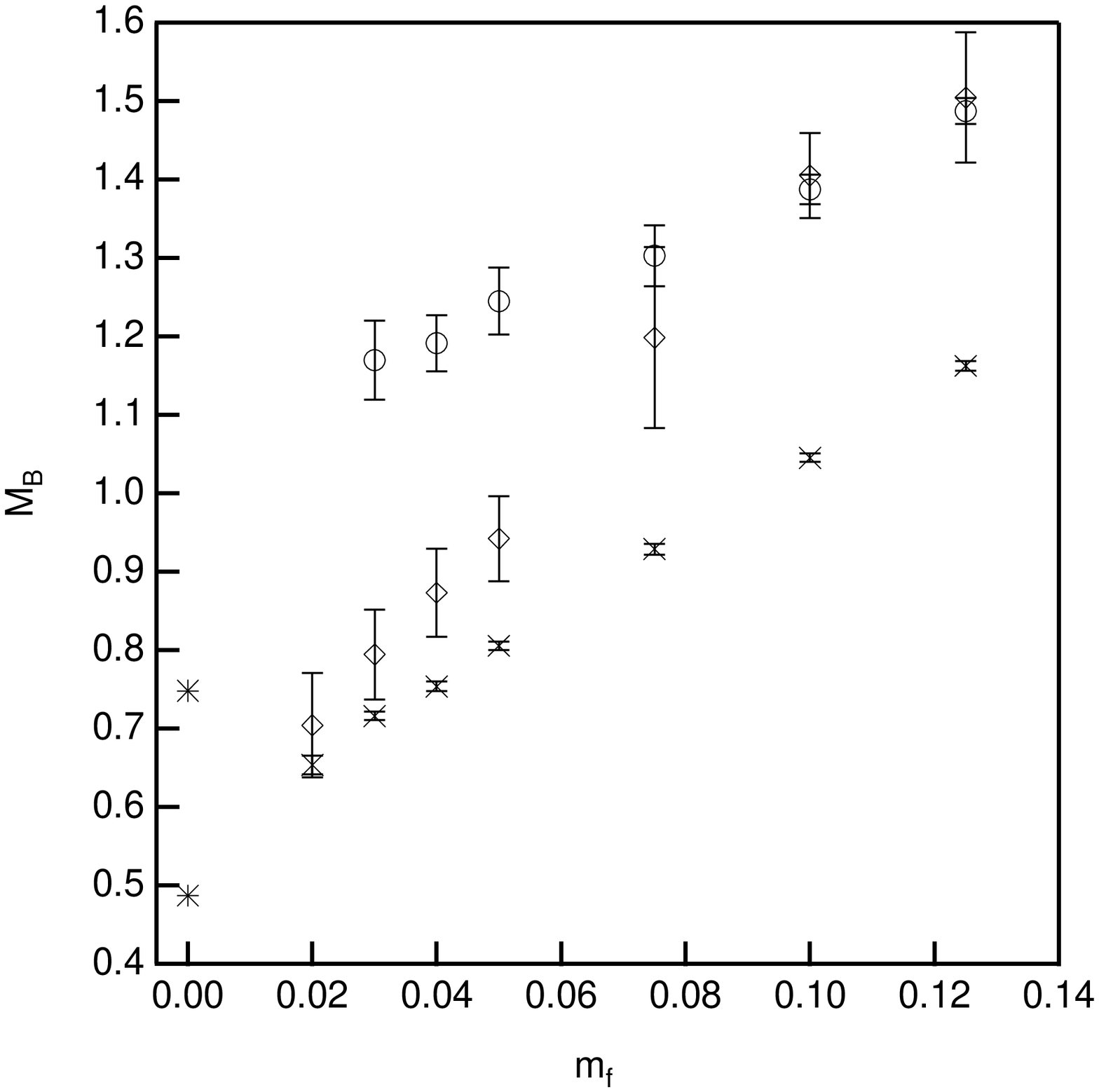}}
\vspace{0.5cm}

\newpage
\centerline{\Large FIG.6 (Phys.Rev.D) Shoichi Sasaki \etal}

\vspace{2.5cm}
%
%
\noindent
\centerline{\epsfxsize=15.0cm
\epsfbox{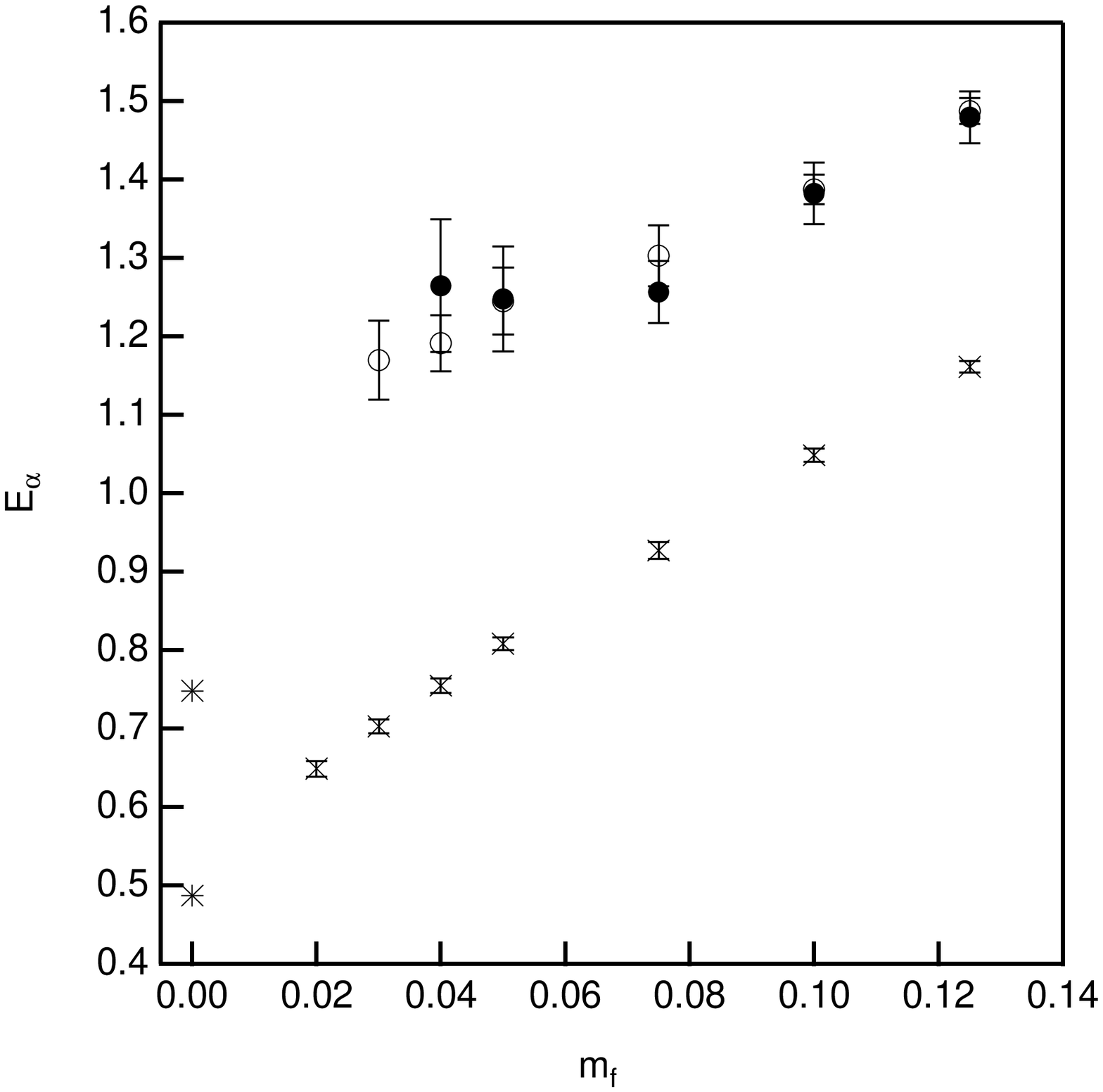}}
\vspace{0.5cm}

\newpage
\centerline{\Large FIG.7 (Phys.Rev.D) Shoichi Sasaki \etal}

\vspace{2.5cm}
%
%
\noindent
\centerline{\epsfxsize=15.0cm
\epsfbox{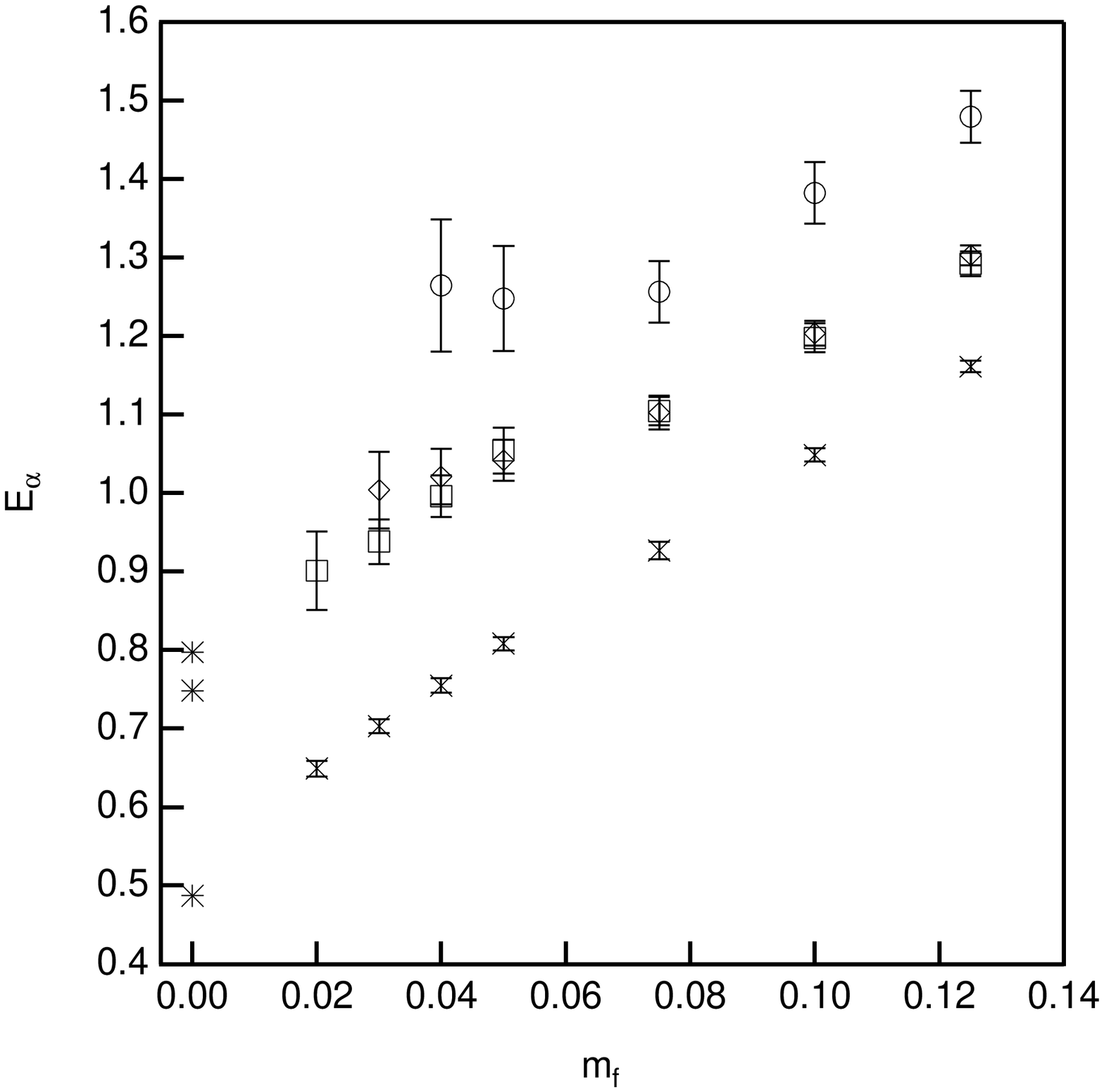}}
\vspace{0.5cm}


%


\end{document}